\documentclass[fleqn,usenatbib]{mnras}

\usepackage{newtxtext,newtxmath}

\usepackage[T1]{fontenc}

\DeclareRobustCommand{\VAN}[3]{#2}
\let\VANthebibliography\thebibliography
\def\thebibliography{\DeclareRobustCommand{\VAN}[3]{##3}\VANthebibliography}

\usepackage{graphicx}	
\usepackage{amsmath}    
\usepackage{hhline}
\usepackage{booktabs}
\usepackage{float}
\usepackage{soul}
\usepackage{longtable}
\usepackage{blindtext}
\usepackage{color}
\usepackage{caption}
\usepackage{hyperref}
\floatstyle{plaintop}
\restylefloat{table}

\usepackage{xcolor}
\newcommand{\nus}{\textit{NuSTAR}}
\newcommand{\xmm}{\textit{XMM-Newton}}

\newcommand{\integral}{\textit{INTEGRAL}}
\newcommand{\swi}{\textit{Swift-BAT}}
\newcommand{\suz}{\textit{Suzaku}}
\newcommand{\bsw}{\textit{BeppoSAX-WFC}}
\newcommand{\igra}{IGR J16320-4751}
\newcommand{\igrb}{IGR J16479-4514}
\newcommand{\igrc}{IGR J16493-4348}
\newcommand{\igrd}{IGR J16393-4643}
\newcommand{\igre}{IGR J17544-2619}

\newcommand{\hma}{4U 1538-52}
\newcommand{\hmb}{4U 1907+09}
\newcommand{\hmc}{Vela X-1}
\newcommand{\hmd}{4U 1700-37}
\newcommand{\hme}{2S 0114+650}
\newcommand{\hmf}{4U 1909+07}
\newcommand{\bhb}{4U 1630-47}
\newcommand{\gs}{GS 0834-430}

\newcommand{\sm}{SMC X-1}
\newcommand{\lm}{LMC X-4}
\newcommand{\oao}{OAO 1657-415}
\newcommand{\gx}{GX 301-2}

\title[ \nus\ study of flaring]{\nus\ investigation of X-ray variability and hard X-ray spectral properties in \igra\ and \igrb}

\author[Varun et al.]{
Varun,$^{1}$\thanks{E-mail: varunaries@aries.res.in}
Gayathri, Raman$^{2}$
\\
$^{1}$Department of Astronomy and Astronomy, Aryabhatta Research Institute of Observational Sciences (ARIES), Manora Peak, Nainital-263001, India\\
$^{2}$Department of Astronomy and Astronomy, The Pennsylvania State University, 525 Davey Lab, University Park, PA 16802, USA\\}

\date{Accepted XXX. Received YYY; in original form ZZZ}

\pubyear{2022}

\begin{document}
\label{firstpage}
\pagerange{\pageref{firstpage}--\pageref{lastpage}}
\maketitle

\begin{abstract}

We present the results obtained from a comprehensive timing and spectral study of two high mass X-ray binary sources using \nus\ observations. These two sources, \igra\ and \igrb\, were discovered by \integral\ and have been characterizated for the first time in the hard X-ray band (beyond 10~keV) in this work. In these sources, we observe the occurrence of intense X-ray flares, with average luminosities exceeding 10$^{36}$~erg~s$^{-1}$. Our analysis reveals that these flares can be described consistently in the quasi-spherical accretion regime. The orbital phase of the first flare in \nus\ observation of \igrb\ matches with the orbital phases of previous flares ($\phi=0.35$) in this source detected by other telescopes. We conclude that this flare occurs as a result of the periastron passage of the neutron star, rather than due to the presence of a corotating interaction region (CIR). Furthermore, from the energy-resolved pulse profile analysis of \igra, we find that the pulse fraction is lower in hard X-rays compared to the soft X-rays. We present the hard X-ray spectral parameters of these two sources using several standard spectral model components. We do not detect a cyclotron absorption feature in either target. We provide estimates to the surface magnetic field strength of NS in \igra\ using two indirect methods. Lastly, we observe spectral hardening during flaring segments compared to the off-flaring segments which indicates that comptonization is more effective during the flaring segments.

\end{abstract}




\section{Introduction}

High mass X-ray binaries (HMXBs) are astronomical sources hosting a neutron star (in most cases) accreting from an O-B supergiant star. For majority of sources, X-ray emission is caused by the stellar wind from the companion star while a handful of sources undergo disk accretion via Roche-lob overflow \citep{2019NewAR..8601546K}. In the wind accreting systems, the accreting matter reaches the neutron star (NS) in time comparable to the free-fall time scale after crossing its accretion radius. Hence, the clumpiness in stellar winds of the companion star is reflected in the variability of the X-ray emission from these sources \citep{1973ApJ...179..585D,2012MNRAS.421.2820O,2016A&A...589A.102B}. The classical supergiant HMXBs (SgHMXB) are persistent sources with a moderate luminosity of $10^{36}$-$10^{37}$ erg $s^{-1}$. In addition, their X-ray emission is characterized by the presence of flares with a duration of hundreds to thousands of seconds and a dynamic range of upto $\sim$100. Flaring activity in SgHMXB sources like \hmc\ and \hmd\ have been used to estimate the physical parameters of stellar wind clumps by including orbital parameters of sources and different impact parameters for clumps \citep{2009MNRAS.398.2152D}. 

A new subclass of HMXBs was discovered by the \integral\ telescope. This subclass, known as Supergiant fast X-ray transients (SFXTs), exhibits characteristic bright flares with dynamic range of $10^2$-$10^6$ \citep{2006ApJ...646..452S,2015JHEAp...7..126R,2017A&A...608A.128B}. In between flares, these sources appear to be less luminous compared to SgHMXB sources with $L_{X}\sim10^{32}$-$10^{33}$ erg $s^{-1}$. Currently, there are about 11 sources that belong to this subclass \citep{2022arXiv221205083R}. X-ray variability studies in SFXTs indicate unrealistically large clump masses on account of the underlying assumption of direct accretion from a clumpy wind \citep{2007A&A...476..335W}. This necessitates additional physical mechanisms to fully explain their behavior.  Some authors have proposed that  centrifugal inhibition of accretion achieves a reduction in X-ray luminosity by a factor of 100-1000 \citep{2007AstL...33..149G}. This is possible in some cases when the NS in SFXTs is in a supersonic propeller regime and encounters with a dense clump which triggers a switch to direct accretion. Alternatively, these systems might contain a young NS with a strong magnetic field ($\mu_{30} \simeq 10-100 $) and a very long spin period ($>$1000 sec). The magnetosphere of NS then inhibits the accretion of material \citep{2008ApJ...683.1031B}. Another possibility is that these sources could be in the settling accretion regime and the collapse of a hot shell onto NS due to a reconnection event could manifest as a bright flare \citep{2014MNRAS.442.2325S}. In this paper, we discuss the flaring activity observed in two HMXB sources \igra\ (an SgHMXB) and \igrb\ (an SFXT) and characterize their spectral and timing properties using  broadband \nus\ observations.

\igra\ was detected by \integral\ as a hard X-ray source, serendipitously in February 2003, during ToO observations of the black hole binary  \bhb\ \citep{2003IAUC.8076....1T}. Archival data from \bsw\ instrument during the period 1996--2002 revealed its persistent X-ray nature \citep{2003IAUC.8077....2I}. The compact object in this source is a NS with spin period $P_{spin}=1309\pm40$ s \citep{2005A&A...433L..41L} and the companion star is identified as a BN0.5 Ia supergiant star \citep{2013A&A...560A.108C}. This source shows a single peak pulse profile with a constant pulse fraction of upto 20 keV. X-ray time series from the \xmm\ data shows variability in terms of flares, the disappearance of pulsation, and changing shape of pulse profile even from observations that were carried out on consecutive days \citep{2023NewA...9801942V}. 

The spectral continuum of \igra\ (using combined data from  \xmm\ and \integral) has been previously described by a power law  ($\Gamma=0.2^{+0.02}_{-0.12}$) with a high energy cut-off ($E_{cut}=7.1\pm0.4$~keV). Alternatively, a thermal comptonization model (with seed photon temperature $kT=1.98^{+0.15}_{-0.10}$~keV and electron temperature $kT_e=8.0^{+1.0}_{-0.7}$~keV) was also known to fit the same data well \citep{2006MNRAS.366..274R}. In addition, the spectrum is characterized by a strong Fe K$_{\alpha}$ line and large absorption column density $N_{H}\sim10^{23}$~cm$^{-2}$. The absorption column density and flux of Fe k$_{\alpha}$ line emission show large variations along its orbit and were found to be correlated with each other. The spectrum below 3~keV shows an excess emission in multiple \xmm\ observations \citep{2023NewA...9801942V}.

\igrb\ was discovered as a hard X-ray transient by \integral\ during the galactic center observations in 2003 \citep{2003ATel..176....1M}. It was classified as an SFXT after the detection of multiple flares \citep{2005A&A...444..221S,2006ApJ...646..452S} and the identification of the companion star as an OB type super-giant \citep{2008A&A...484..783C}. The best estimates of source distance lie in the range of 4.4--4.6~kpc \citep{2015ApJ...808..140C}. This binary has an orbital period of $\sim$3.3 ~days \citep{2009MNRAS.397L..11J} during which an eclipse of 0.6~days is also observed. Additionally, super-orbital modulation of $\sim$11.88 days has been observed using \integral\ and \swi\ data \citep{2013ApJ...778...45C}. More than 20 strong flares have been detected in \igrb\ with \integral, \swi, and \suz\ observatories \citep{2008A&A...487..619S,2009A&A...502...21B,2013MNRAS.429.2763S}.

An \xmm\ observation of \igrb\ in 2008 and 2012 covered the early phases of its eclipse (ingress and a small portion of the eclipse). X-ray spectra from these observations were fitted with the spectral model consisting of three power-laws, two Gaussian components and two photoelectric absorption components \citep{2008ApJ...683.1031B,2020ApJ...900...22S}. The three power-law components were interpreted as  1) direct emission from the compact object, 2) scattered emission from stellar wind, and 3) emission from dust scattering. Alternatively, the eclipse spectrum of this source can be fitted with heavily absorbed power-law, two Gaussian components, and less absorbed blackbody components \citep{2019ApJS..243...29A}. The out-of-eclipse spectrum observed with \suz\/-XIS instrument (0.2--10 ~keV) is well described by a power-law ($\Gamma\sim1.3$) with a large column density ($N_{H}\sim10^{23}$ cm$^{-2}$) and a narrow emission line at 6.4~keV ($EW=70\pm10$ eV) \citep{2013MNRAS.429.2763S}.

Broadband X-ray spectral characteristics of both these targets are inadequately understood owing to limited high-energy observations beyond 10 keV. With high timing accuracy and moderate spectral resolution in the 3-79 keV band, \nus\ has significant potential for investigating the hard X-ray properties of accreting HMXBs. In this paper, we analyze archival observations for these two targets, with the objective of investigating the hard X-ray properties and comprehending their temporal variability, particularly through an examination of the captured flaring activity.
The manuscript is organized as follows. Section \ref{sec:obda} provides details about the observation and data reduction  methods. Section \ref{sec:tian} describes the temporal data analysis procedure and results. In section \ref{sec:span} and \ref{sec:trsp}, we present the details of time averaged and time resolved spectral analysis. Finally, results from the timing and spectral analysis are discussed in section \ref{sec:dis}.

\section{Observations and data reduction} \label{sec:obda}

\nus\ observed \igra\ in June 2015 and December 2018. Two observations of \igrb\ were were carried out in April 2019, separated by $\sim$10 days. The details of these observations are given in Table \ref{tab:obs}.  \nus\ comprises of two identical but independent telescopes containing hard X-ray focusing optics and cadmium zinc telluride (CZT) pixel detectors. X-ray optics modules of \nus\ consists of 133 multilayer grazing incidence shells which enable a significant collecting area upto 79 keV \citep{2013ApJ...770..103H}. Both the focal plane modules consist of a two-by-two array of detectors, each with an array of 32x32, 0.6~mm pixels. In order to avoid pileup for bright sources (as seen in many CCD based X-ray detectors) in this instrument, the pixels are read out individually upon triggering. This enables each \nus\ module to handle count rates of upto $\sim$400 ct s$^{-1}$ with no pileup problem. A combination of large collecting area, high resolution, and timing accuracy makes \nus\ ideally suited to study X-ray binary sources.

The raw event files were passed through 2-stage processing using NUSTARDAS pipeline v2.1.1 with CALDB version 20220802. The first stage involves data calibration, cleaning using "good time intervals" (GTIs), and screening for passage through South Atlantic Anomaly (SAA). In the second stage, user products (image, light curve, and spectrum) are created from event files. Source counts are taken from a circular region with a $60\arcsec$ radius. A nearby source free circular region of 90$\arcsec$ is chosen for the background.  Light curves were barycenter corrected with FTOOL {\fontfamily{qcr}\selectfont barycorr} using DE-200 solar system ephemeris. The scientific data analysis of user products have been carried out using HEASOFT V6.29.

\section{Timing analysis} \label{sec:tian}

\begin{figure}
    \centering
    \includegraphics[trim= 1.5cm 11.5cm 0cm 1.5cm, scale=0.37,angle=0]{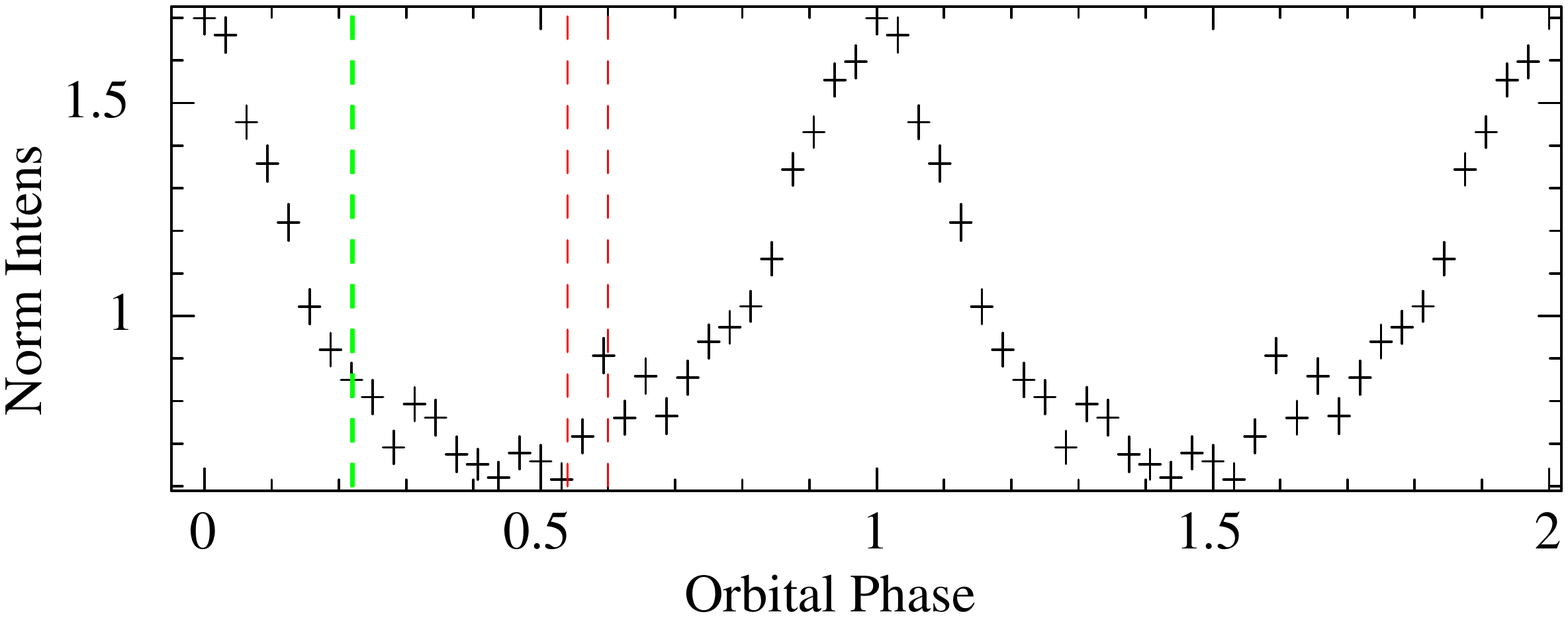}
    \includegraphics[trim= 1.5cm 10cm 0cm 0cm, scale=0.37,angle=0]{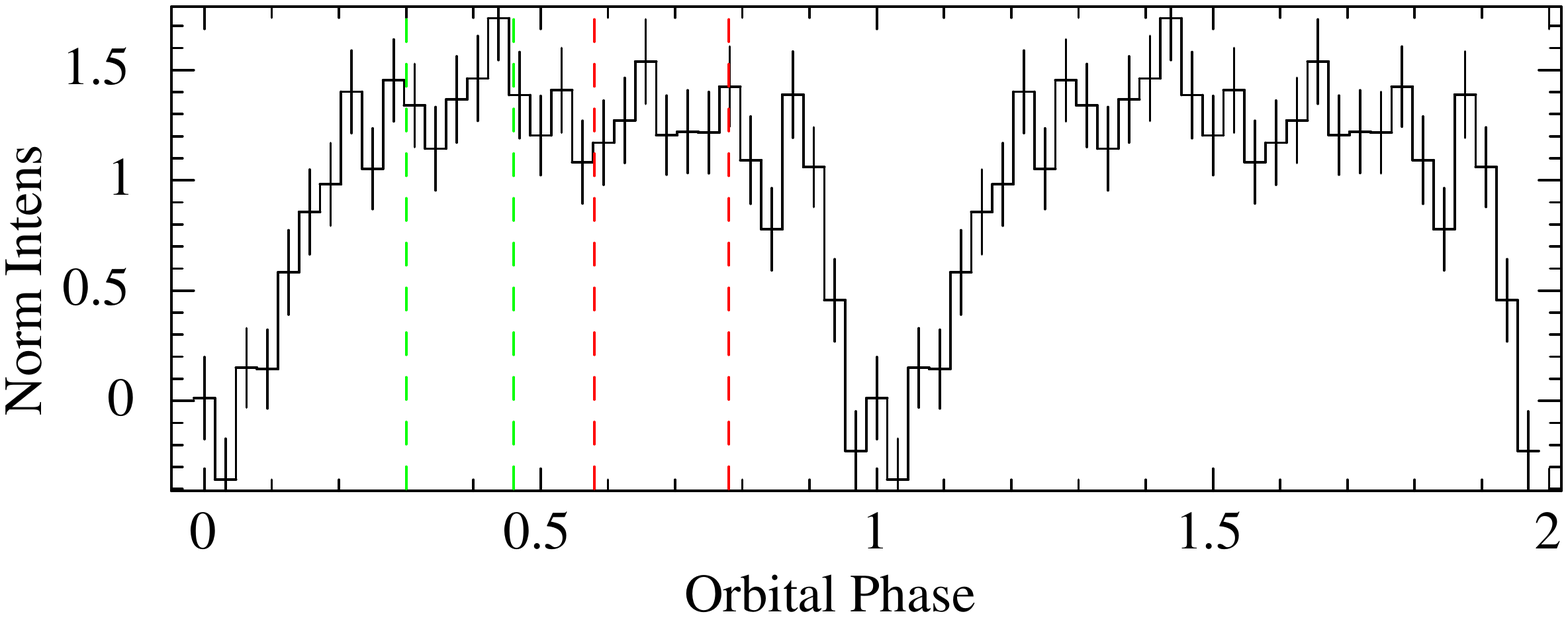}
    \caption{Orbital profiles of \igra\ (top) and \igrb\ (bottom) created from \swi\ data (15-50 keV) between 2005-2022. The full orbit is divided into 32 phase bins for both sources. Vertical dashed lines in green and red marks the orbital phases of the \nus\ observations.}
    \label{fig:opa}
\end{figure}

Figure \ref{fig:opa} shows the orbital intensity profiles of \igra\ and \igrb, created from \swi\ \citep{2013ApJS..209...14K} light curves in the 15--50~keV band.  The orbital profiles were generated by folding the light curve at an orbital period of P$_{orb}$ = 8.991 days \citep{2023NewA...9801942V} and 3.31961 days \citep{2015ApJ...808..140C}, with a reference epoch of MJD 53417 and MJD 53415, respectively, for \igra\ and \igrb. The folded orbital profile of \igra\ is similar to \hmb\ and \hme\ showing a large variation in intensity with a broad maximum and minimum. The change in orbital intensity can be explained by a simple model consisting of a neutron star orbiting an OB supergiant embedded in its dense stellar wind whose profile is given by CAK model \citep{2018A&A...618A..61G}. The orbital intensity profile of \igrb\ is typical of an eclipsing binary with high intensity during most part of orbital cycle and a deep minimum (with orbital phase span $\sim$0.25) during which the intensity falls nearly to zero. Both \nus\ observations of \igra\ are taken close to the minima in its orbital profile. The two \nus\ observations of \igrb\ are taken out of eclipse where its orbital intensity is high.

\begin{figure}
    \centering
    \includegraphics[trim=1.5cm 14cm 0cm 1.5cm, scale=0.37,angle=0]{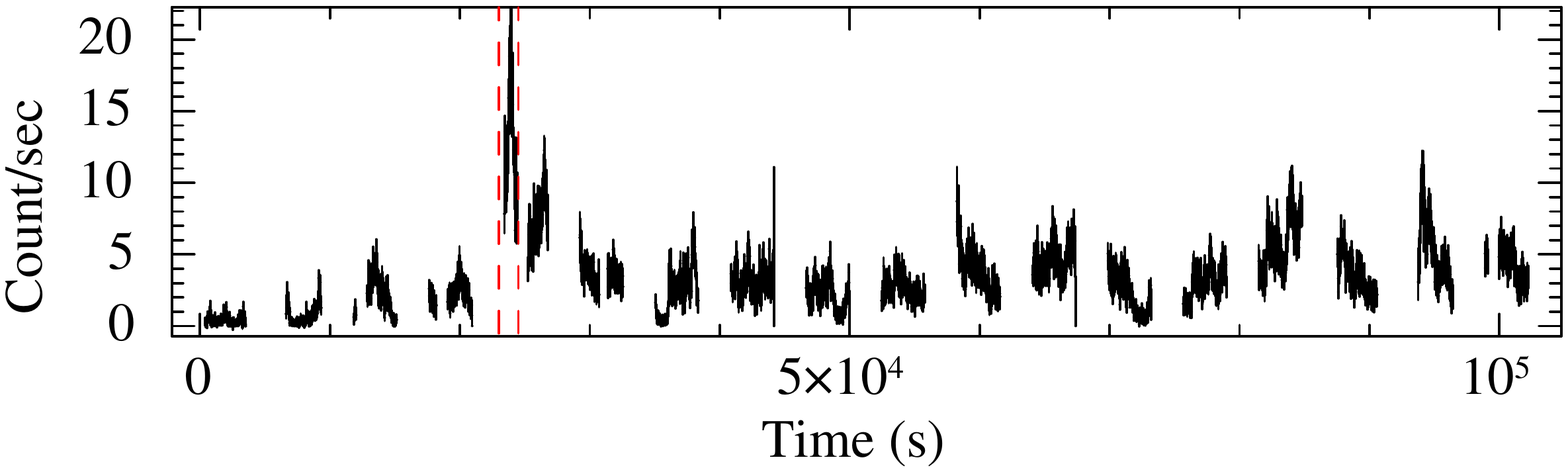}
    \includegraphics[trim=1.5cm 14cm 0cm 0cm, scale=0.37,angle=0]{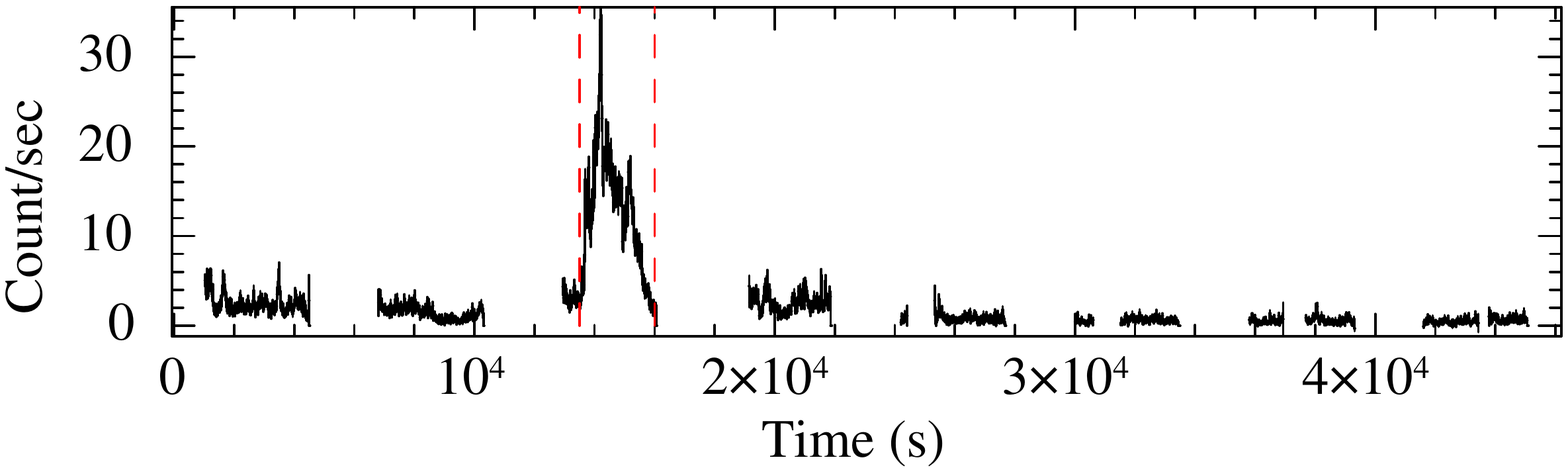}
    \includegraphics[trim=1.5cm 13cm 0cm 0cm, scale=0.37,angle=0]{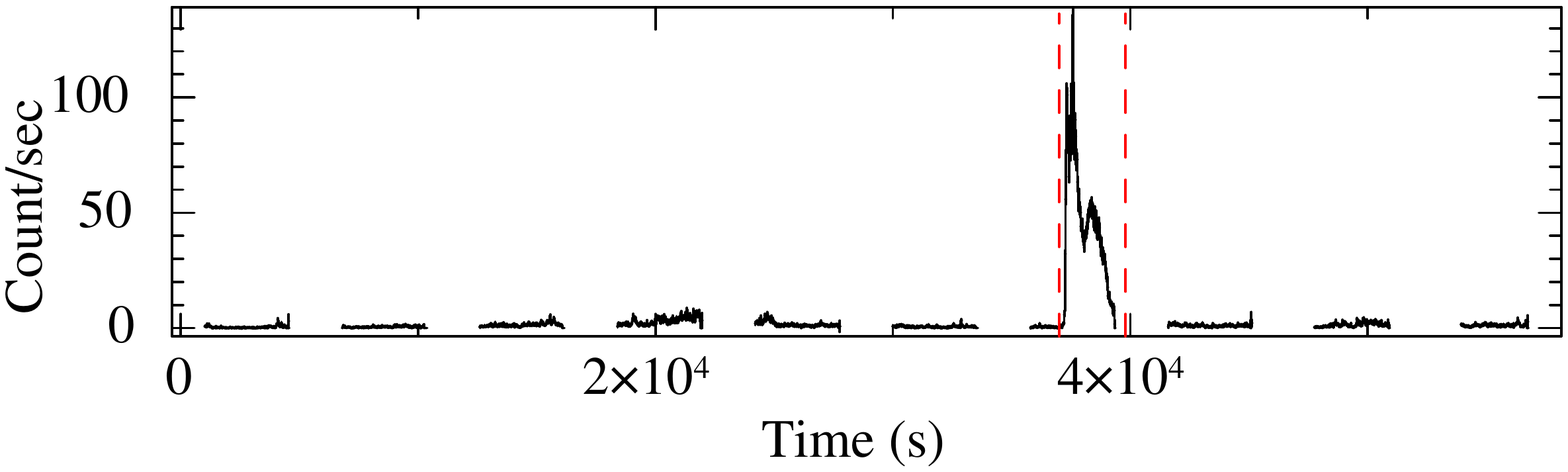}
    \caption{ Top panel shows the \nus-FPMA light curve of \igra\ with time bin size of 20 s in 3-79~keV band. The following two panels (middle and bottom) correspond to the light curves from the two observations of \igrb. Red dashed lines in all three panels are used to indicate the flaring activity.}
    \label{fig:lcs}
\end{figure}

\begin{figure}
    \centering
    \includegraphics[trim= 1.5cm 1.5cm 0cm 1.5cm, scale=0.36,angle=0]{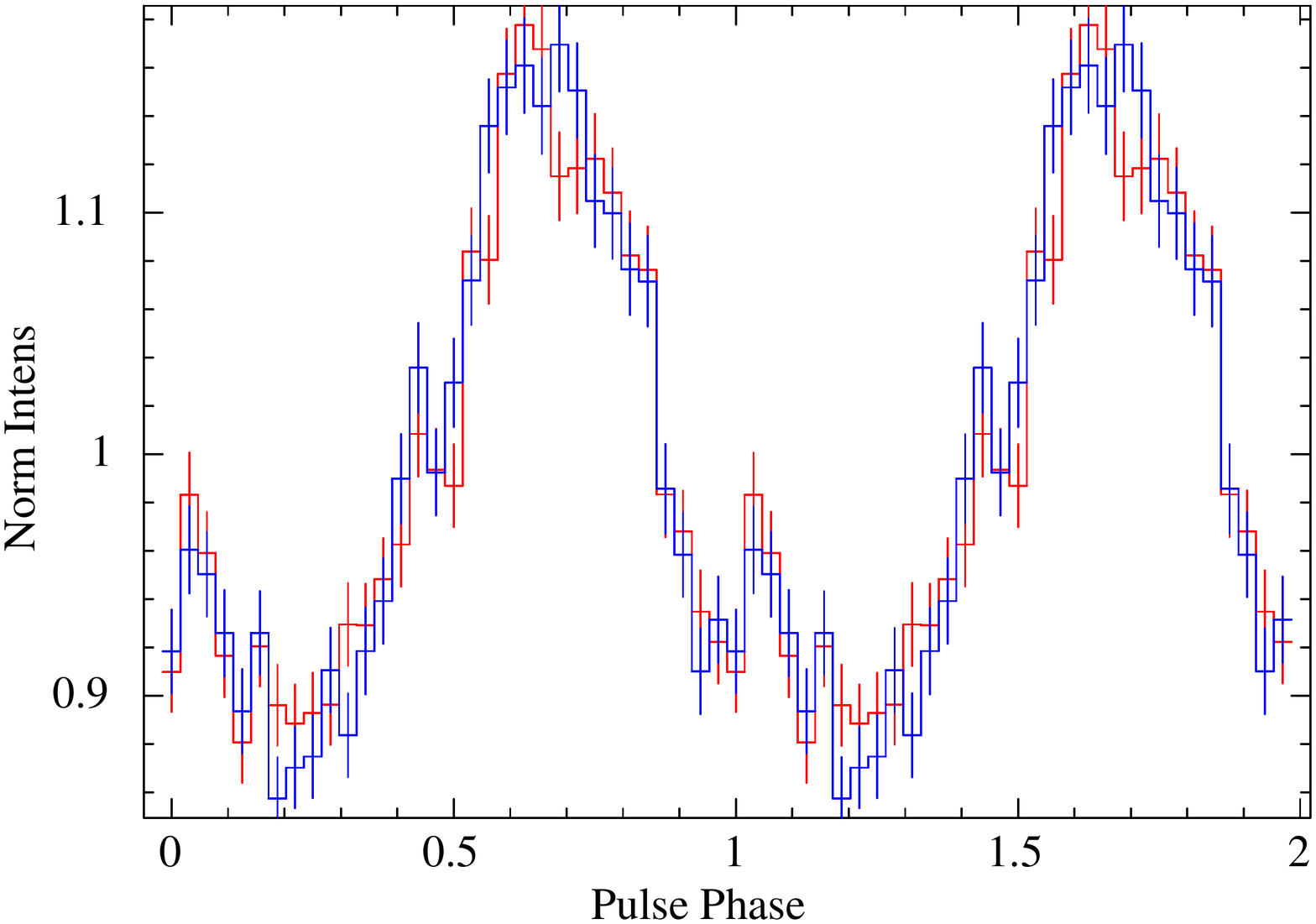}
    \caption{Time averaged pulse profile of \igra\ created from \nus\ light curves folded at the spin period of 1306~s and 32 phase bins. Red and blue data points correspond to modules FPMA and FPMB respectively. }
    \label{fig:ppf}
\end{figure}

\begin{figure}
    \centering
    \includegraphics[trim=5.2cm 0cm 3cm 2.2cm, scale=0.55,angle=0]{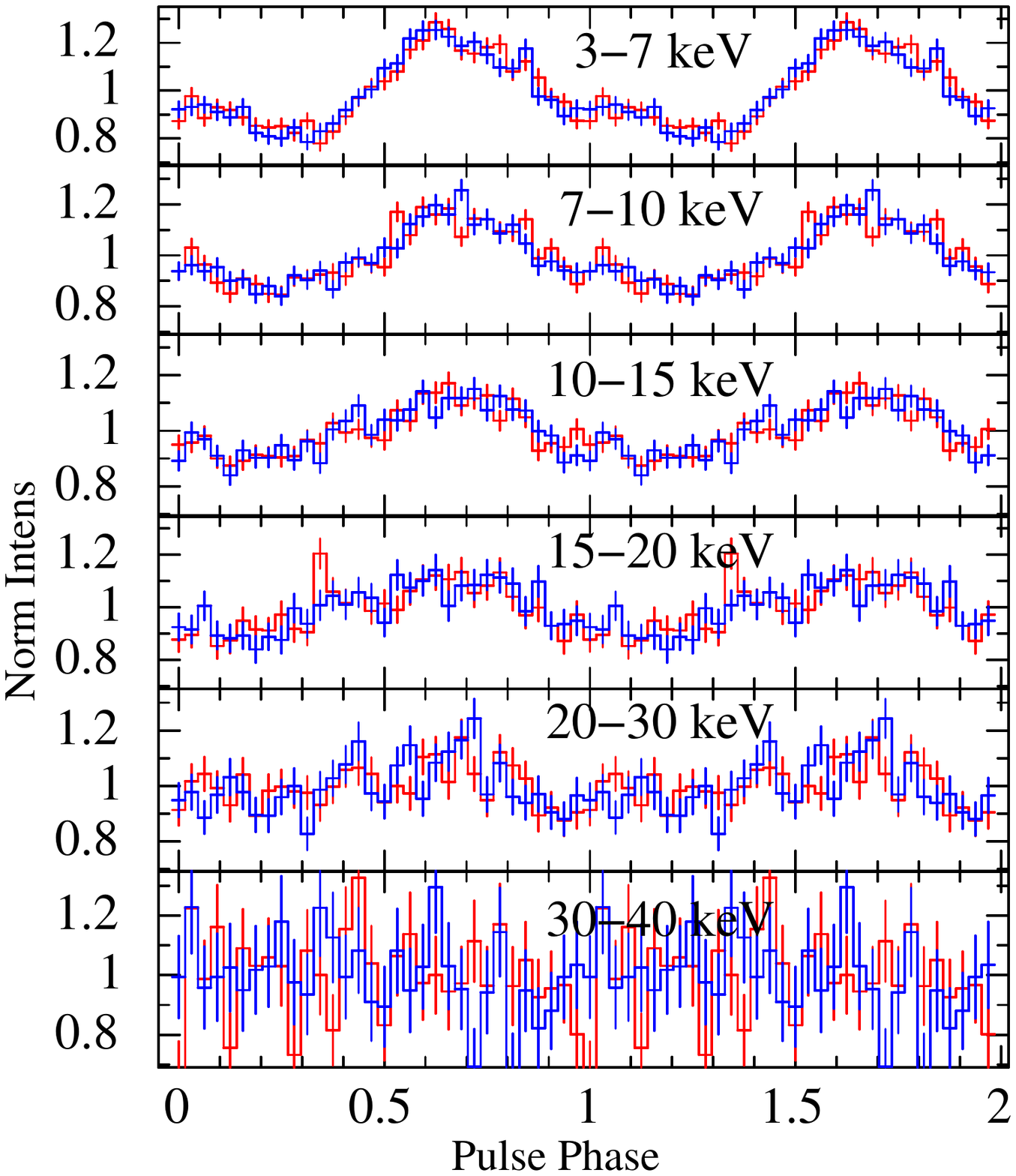}
    \includegraphics[trim=0.5cm 0.5cm 2cm 1.1cm, scale=0.45,angle=0]{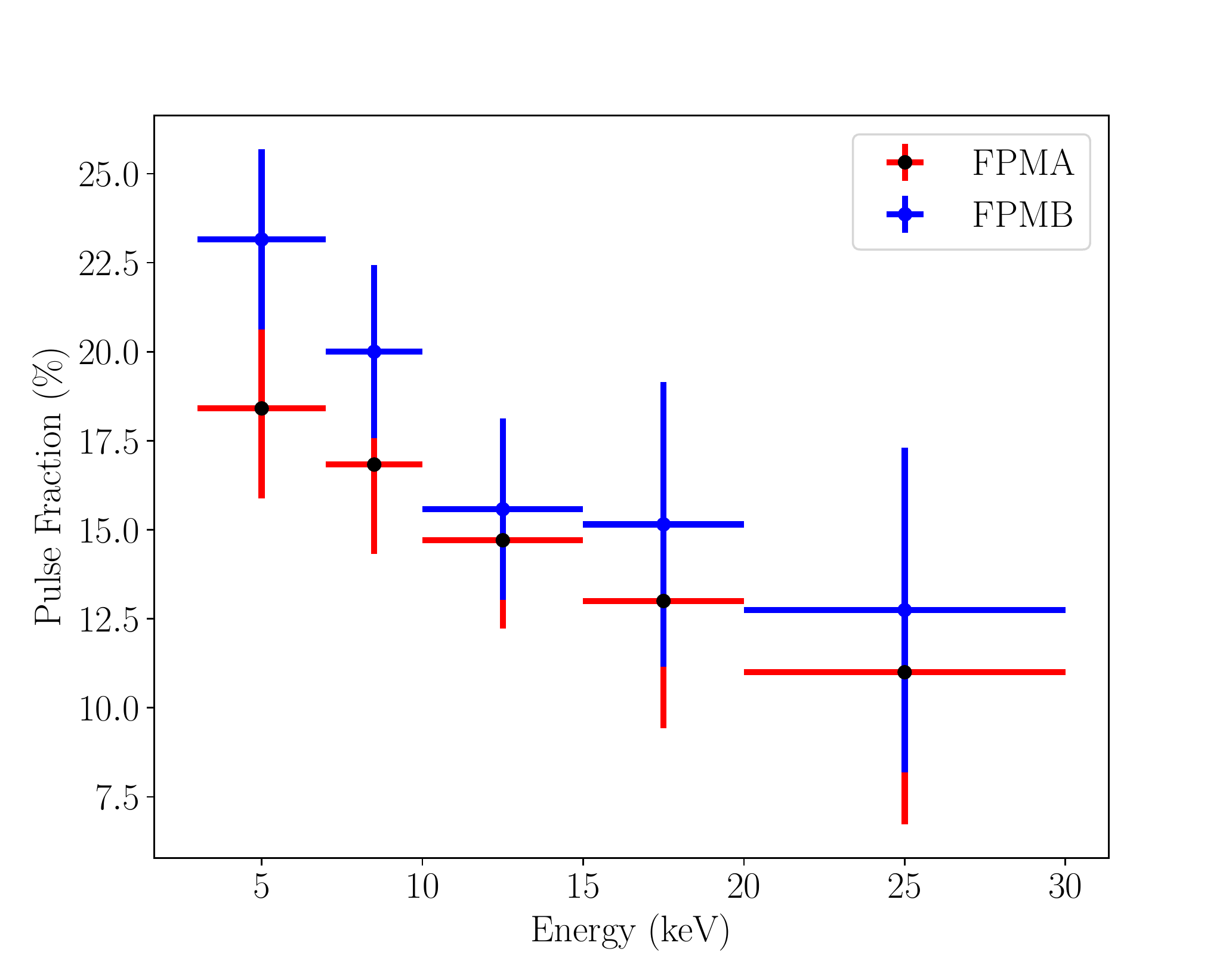}
    \caption{Top panel shows the energy resolved pulse profiles of \igra\ in six energy bands. Red and blue data points are from FPMA and FPMB modules. Bottom panel shows the variation of pulse fraction in the main peak of pulse profile as a function of energy. Central value of each energy band represents the whole band and width of band is taken as X-error. }
    \label{fig:pper}
\end{figure}

We extracted the \nus\ light curve of \igra\ with 1s time resolution in the full 3-79 keV band. The light curve from the first observation which has a duration of $\sim$102 ks is shown in Figure \ref{fig:lcs} with time bin size of 20 s.  Flaring activity is observed in this observation during which the count rate increases by a factor of $\sim$5 (marked with two red dashed lines). In addition, short term count rate variations are also observed, as reported in the previous works for this source \citep{2018A&A...618A..61G}. The second observation of \igra\ contains 4 segments ($\sim$500 s) and does not contain any flare. We searched for periodicities in the first observation with {\fontfamily{qcr}\selectfont efsearch} tool which uses epoch folding technique \citep{1983ApJ...266..160L} and found a period value of $P_{spin} = 1306\pm14$ s. The uncertainty on the pulse period value was determined by carrying out a bootstrap procedure on a set of thousand Gaussian randomized light curves \citep{2013AstL...39..375B}. This spin period value is in agreement with the previously reported value of this source from \xmm\ observations \citep{2006MNRAS.366..274R,2023NewA...9801942V}. We folded the light curve from the first observation with the best spin period value and 57179 MJD as reference epoch to obtain the pulse profile in full energy band. The pulse profile contains a broad single peak followed by a minima and a small interpulse in between them (see \ref{fig:ppf}). Pulse profile (see Figure \ref{fig:ppf}) contains a broad single peak followed by a minima and a small interpulse in between them. This is very similar to the pulse profile seen in the soft energy band using \xmm\ data. Based on the good count rate of this source and the sensitivity of \nus, we decided to extract the data in several energy bands: 3--7, 7--10, 10--15, 15--20, 20--30, 30-40, and 40-50~keV. We searched for the spin period in these light curves and found that the pulsation is detected only upto 30 keV. The energy resolved pulse profiles obtained by folding these light curves using spin period determined from the 3-79 keV NuSTAR light curve, are shown in the top panel of Figure \ref{fig:pper}. The shape of pulse profile in different energy bands is similar to the full energy band pulse profile. The pulse peak becomes broader in harder energy bands compared to the softer energy bands. The pulse profile shape remains more or less unchanged in the different energy bands, while the pulse fraction is observed to decrease from 22\% in the 3--7~keV band to 10\% in the 20--30~keV band as shown in the bottom panel of Figure \ref{fig:pper}. 

Similarly, For \igrb\ also, we extracted light curves with 1~s time resolution in the full 3--79 keV band. The light curves from the two observations are shown in the middle and bottom panels of Figure \ref{fig:lcs}, respectively with the time bin size of 20~s. The count rate in both observations is low ($<$5 ct s$^{-1}$) with some variability between different orbit data. Both observations contain one flare each with a duration of 2700~s (in observation 1) and 2100~s (in observation 2). The first flare has a slow rise and slower decline with a time scale of 700~s and 1800~s respectively. Comparatively, the counts during the second flare rise quickly to $>$100 in 100~s. The peak count rate during the flare in the first observation is 35~ct~s$^{-1}$ whereas the flare in the second observation is stronger with a peak count rate of 135~ct~s$^{-1}$. We searched for the periodicity in the light curved of both observations using {\fontfamily{qcr}\selectfont LombScargle} class of {\fontfamily{qcr}\selectfont Astropy} package \citep{2012cidu.conf...47V,2015ApJ...812...18V}. The Lomb-scargle periodogram is designed to detect periodic signals in unevenly spaced observations. Periodicities were searched in the frequency range from 0.0002 to 0.15 (Nyquist frequency of data set). No significant and consistent periodicity was observed from both data sets.

\begin{table*}[ht!]
\centering
 \begin{tabular}{||c c c c c c c||} 
 \hline
Source Name     & Obs ID        & Start time   & Mid time    & Stop time     & Exposure     & Orbital     \\
              &                & MJD (day)    & MJD (Day)   &    MJD(day)      & (FPMA/B)(ks) & Phase    \\
\hline\hline
\igra           & 30101026002   & 57179.78202  & 57180.38272 & 57180.98341   &  49.89/49.89 & 0.57$\pm0.03$     \\
                & 90401373002   & 58480.77510  &  58480.88621 & 58480.99732   &  2.32/2.38   & 0.2200$\pm0.0015$    \\ 
\hline
\igrb           & 30402028002   & 58574.79595  & 58575.79595 & 58575.31677   &  21.86/21.79 & 0.38$\pm$0.08    \\
               & 30402028004   & 58582.38274  & 58582.71087 & 58583.03899   &  31.23/31.05 &  0.68$\pm$0.10  \\
\hline 
 \end{tabular}
 \caption{Details of \nus\ observations used in this work.}
 \label{tab:obs}
\end{table*}

\section{Time averaged Spectral analysis} \label{sec:span}

\subsection{Spectral models}

Modeling X-ray spectra accurately for accretion-powered X-ray pulsars is currently a challenge owing to the complex accretion flow processes and the presence of a strongly magnetized neutron star, star, both of which contribute to the emission. A power-law model in its basic functional form in combination with an energy cutoff and smoothing function is used to describe the continuum spectra observed in these sources. Some of the widely used mathematical models for HMXB pulsars are 1) high energy cutoff power-law (HIGHECUT in XSPEC), 2) power-law with high energy exponential roll-off (CUTOFFPL in XSPEC), 3) a power-law with Fermi-Dirac cutoff \citep[FDCUT]{T86}, 4) a smooth high energy cutoff model \citep[NEWHCUT]{burderi2000}, 5) negative and positive power-law exponential model \citep[NPEX]{makishima1999}. These models give straightforward estimates of source luminosity and spectral steepness which can be used for characterizing the variation in the source behavior and comparing the properties with other objects of the same class. Alternatively, some analytical models are also used to model spectra of accretion-powered X-ray pulsars in terms of physical parameters. These models describe the comptonization of soft seed photons in a hot plasma. The following models are commonly used for the spectral modelling of HMXB pulsars: 1) COMPTT \citep{1994ApJ...434..570T}, 2) NTHCOMP \citep{1999MNRAS.309..561Z}, 3) COMPMAG \citep{2012A&A...538A..67F}. \citet[in short ST80]{1980A&A....86..121S}  solved the problem of thermal comptonization of spectra for the non-relativistic electron temperatures and diffusive transport of photons. The COMPTT and NTHCOMP models are based on the works which have some success in extending the ST80 work in the relativistic temperature regime and for low optical depths regions. The COMPMAG model is recently developed specially for the accretion at the polar cap of a neutron star with a magnetic field $\gtrsim$ 10$^{12}$ G. In addition, both thermal and bulk comptonization processes have been considered in this model. For bulk comptonization, the solution of the radiative transfer equation can be obtained for matter flow with velocity increasing towards the neutron star ($\beta(z) \propto z^{-\eta}$, here $\beta=v/c$ is the infalling matter velocity and z is distance from NS surface ) or matter flow stagnating towards the neutron star ($\beta(z) \propto -\tau$, here $\tau$ is the vertical optical depth of accretion column). This model has a total of 9 parameters. While fitting the spectra using this COMPMAG model, we fixed the parameters corresponding to electron velocity profile, accretion column size, and neutron star albedo fixed at appropriate guess values and later varied them separately to check whether any variation in these parameters affects the fit quality. All the empirical and analytical models used in this paper are either included as standard models or imported as local packages in XSPEC \citep[][]{2001ASPC..238..415D}. We have used XSPEC version 12.12.0 for all the spectral fitting in this paper.

\subsection{\igra}

\begin{figure}
    \centering
    \includegraphics[trim= 5cm 0cm 0cm 2.3cm, scale=0.55,angle=0]{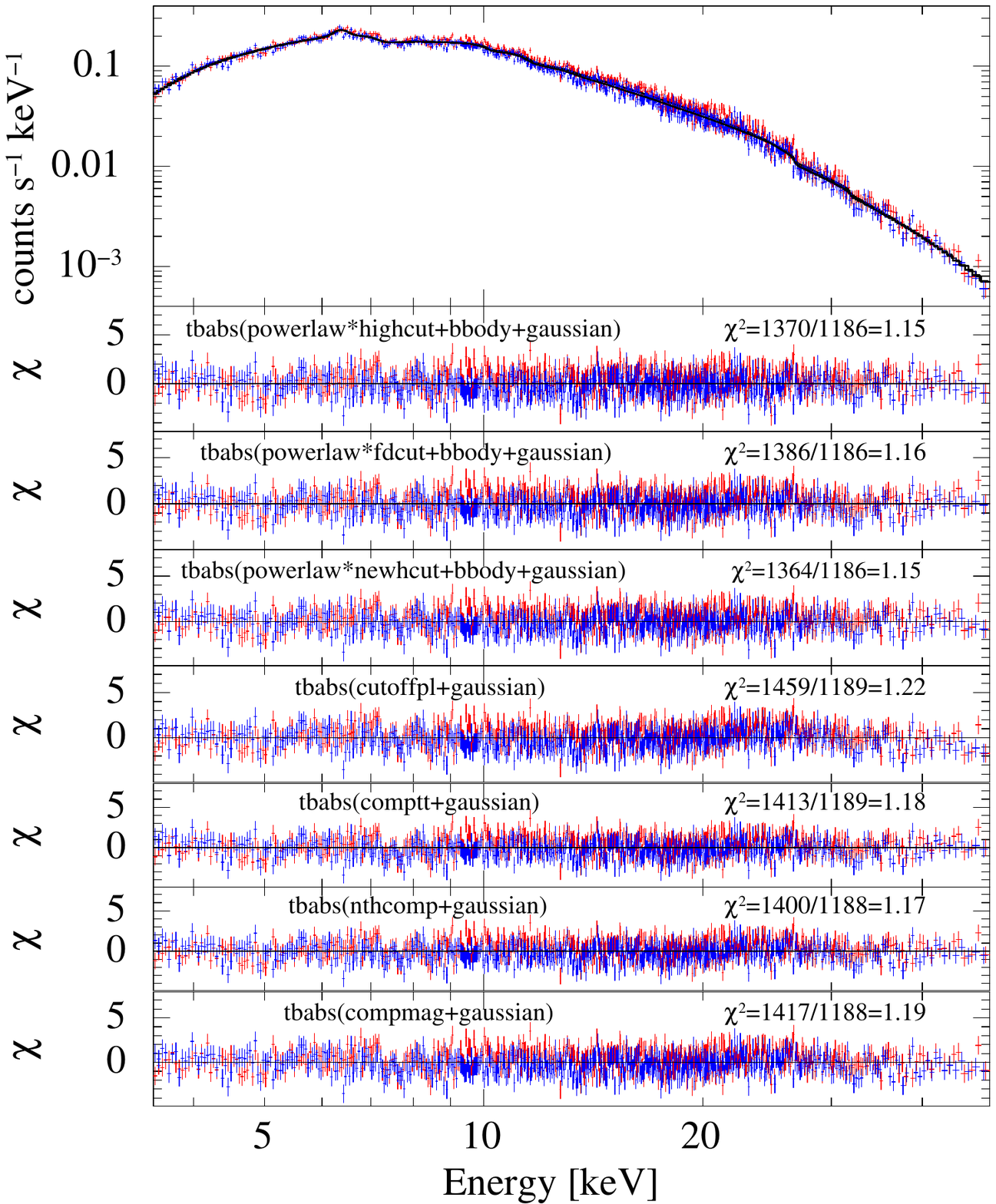}
    \caption{Top panel: Time averaged spectrum of \igra\ with best-fit model. FPMA, FPMB data points are shown with red, blue color whereas black solid curves the total spectral model. Lower panels show residuals of data with respect to different continuum models. }
    \label{fig:spa}
\end{figure}

Spectral data of \igra\ were extracted from all the channels of the focal plane modules in the 3.0--79.0~keV energy band. However, the data statistics were good  in the 3.5–50.0 keV energy range, while beyond 50 keV the spectrum was dominated by background. We rebinned the spectra from Focal Pane Modules (FPMA and FPMB) such that a minimum signal-to-noise ratio of 7 was obtained. We simultaneously fitted the data from both the \nus\ detectors. The background subtracted spectrum was fitted using all the empirical models mentioned above with all the spectral parameters set free. Line of sight absorption was included using {\fontfamily{qcr}\selectfont TBabs} \citep{2000ApJ...542..914W} model. We have used  the {\fontfamily{qcr}\selectfont wilm} abundances table \citep{2000ApJ...542..914W} and the {\fontfamily{qcr}\selectfont vern} cross-section table \citep{1996ApJ...465..487V} for different elements. The fitting with NPEX model resulted in very low normalization for the positive power-law, rendering it identical to the HIGHECUT model. This HMXB has a very prominent emission feature in 6-7 keV energy range \citep{2018A&A...610A..50P,2018A&A...618A..61G,2023NewA...9801942V}. This emission feature is also detected in our current \nus\ data and has been modeled as a single additive Gaussian component. The centroid energy of this emission feature was determined at $\sim$6.4 keV for all the continuum models that were tested. The width of the Gaussian component was fixed at 10$^{-6}$ keV, as it could not be constrained when kept free to vary. Below 5~keV, some additional residuals were seen, especially with the HIGHECUT and NEWHCUT models. A soft blackbody component was found to improve the residuals and the value of fit statistic ($\Delta\chi^{2}=256$ and 263, for HIGHECUT and NEWHCUT, respectively) . In comparison, the partial covering fraction absorption component (PCFABS) was found to be ineffective in this regard. HIGHECUT and NEWHCUT models along with an additive Blackbody component give comparable fits and both better compared to the other remaining two models. When a blackbody component was added to the CUTOFFPL model, its parameter values could not be constrained. The best fit spectral parameters obtained using empirical models are given in Table \ref{tab:pha}. The best fit value of $N_{H}$ and $E_{cut}$ parameters are 21$\times10^{22}$ atoms cm$^{-2}$ and 25 keV, respectively (consistent among different empirical models except the CUTOFFPL model). The blackbody temperature of $\sim$3~keV is a bit higher than usually seen in other accreting compact objects but similar values have been reported earlier with \xmm\ data for this source \citep[][]{2003A&A...407L..41R}.

We also fitted this spectral data with above mentioned physical models. Best fit spectral parameters obtained from different physical models are given in Table \ref{tab:tha}. COMPTT and NTHCOMP models fit the spectrum satisfactorily with $\chi^2/dof$ value of 1413/1188 and 1400/1187, respectively, which is similar to the values obtained with empirical models. The value of seed photon, electron temperature, and optical depth with COMPTT models are 1.84 keV, 9.4 keV and 10.8 keV, respectively. The value of analogous parameters in the NTHCOMP gives similar values. The value of absorption column density $N_H\sim16\times10^{22}$ atoms cm$^{-2}$ with physical models is comparatively smaller than obtained with empirical models. For the spectral analysis using the COMPMAG model, we fixed the accretion column $r_{0}=0.25$ (in units of the NS Schwartzchild radius) and NS albedo A = 1. For the infalling material, we assumed a free-fall velocity profile with the index of velocity profile $\eta=0.5$ and terminal velocity at the NS surface $\beta_{0}=0.05$. The remaining parameters of this spectral model were kept free. A good spectral fit was obtained for this model with $\chi^2/dof=1405/1187$. The value of seed photon and electron temperature obtained with this model (as reported in Table \ref{tab:tha}) are comparatively smaller than the other two physical models. A much smaller value for optical depth, $\tau$ was obtained using this model. However, the optical depth in this model is related to classical optical depth as $\tau\sim10^{-3}\tau_{T}$ \citep{2007ApJ...654..435B}. No significant variation of spectral fit quality or spectral parameters was obtained when $r_{0}$ was changed to 0.5, 0.75, and 0.1. A similar effect was observed for setting $\beta_{0}=0.2$. A plot of spectral fitting and residuals from all the empirical and physical models is shown in Figure \ref{fig:spa}. 

\subsection{\igrb}

\begin{figure}
    \centering
    \includegraphics[trim= 5cm 0.5cm 0cm 3.5cm, scale=0.55,angle=0]{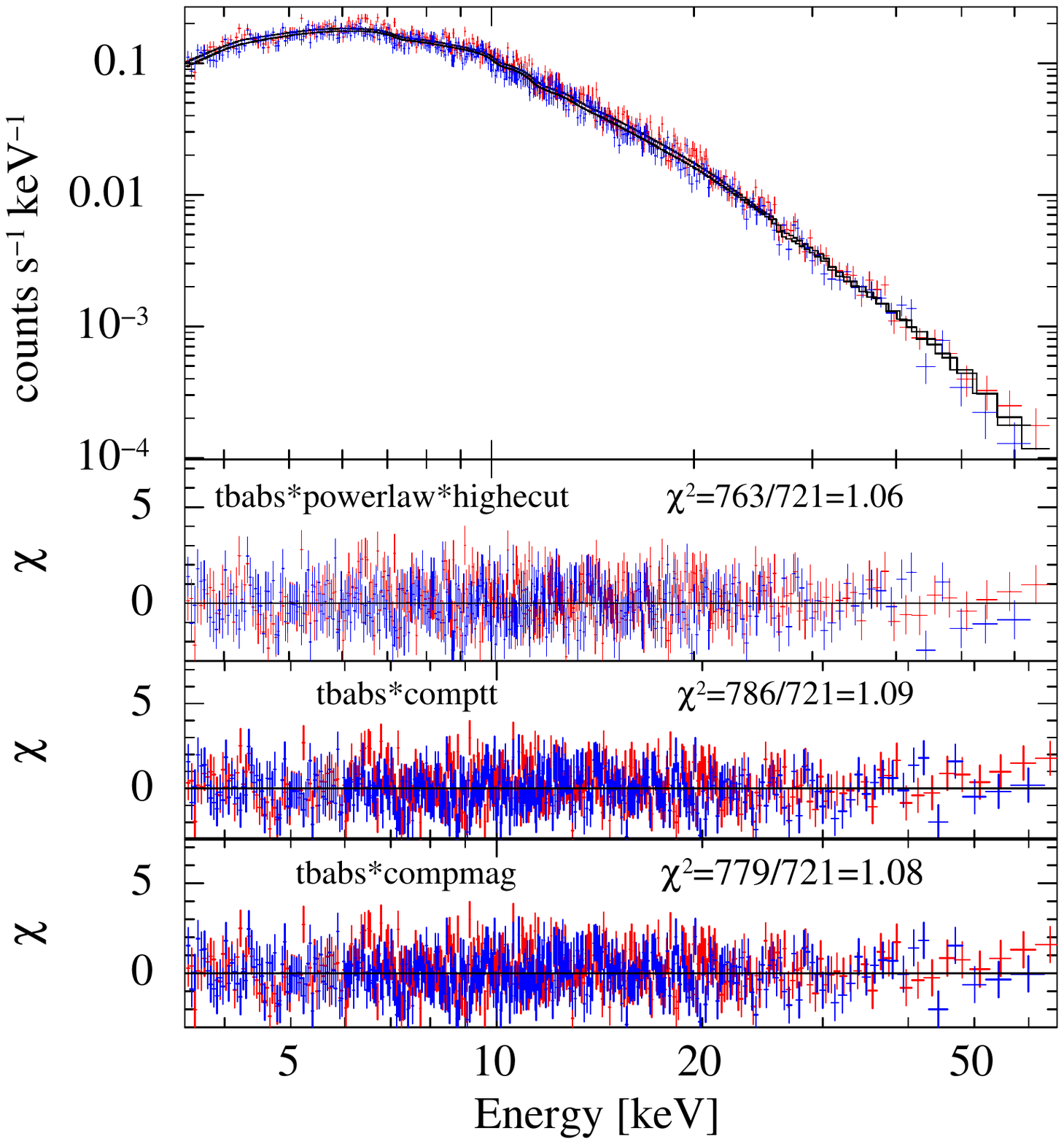}
    \caption{Top Panel: Time averaged spectrum of \igrb\ from the first observation with best-fit model. Red, blue data points and solid black curve represents FPMA, FPMB data, and continuum model respectively. Lower panels shows the residuals of data with respect to HIGHECUT, COMPTT, and COMPMAG models.}
    \label{fig:spb}
\end{figure}

For \igrb, good spectral statistics were available in the 3.5--70.0 keV energy range when the data was rebinned to have a minimum SNR of 7. At present, there are no direct proofs to ascertain the true nature of the X-ray source in this system. However, the spectral properties as determined from previous studies indicate that the compact object in this system might be a neutron star. So, we used three pulsar-like spectral models for this source: one empirical model-HIGHECUT, one commonly used physical model-COMPTT, and the most suitable model for the physical conditions of accretion-powered pulsars-COMPMAG. Among all three models, the HIGHECUT model fits the data best for observations of this source. However, the fitting quality degrades for the second observation with all the spectral models. The spectrum of \igrb\ from observation 1 fitted with the HIGHECUT model and residual from all three models are shown in Figure \ref{fig:spb}. The parameter values obtained from the fitting of data from both observations are given in Table \ref{tab:phbo}. The value of column density increases from 9$\times10^{22}$ atoms cm$^{-2}$ during the first observation to $13\times10^{22}$ atoms cm$^{-2}$ in the second observation. Similarly, there is an increase in spectral index $\Gamma$ from 1.16 to 1.36. The value of cutoff energy $E_{cut}$ and folding energy $E_{fold}$ are consistent within errors from both observations. The value of column density using physical models is smaller compared to empirical models but the trend observed during the two observations are similar in nature. The value of seed photon and electron temperatures using COMPTT model is $\sim$1.5 keV and $\sim$10 keV respectively (consistent between observations). The optical depth of the comptonizing medium decreases slightly from 8.0 during the first observation 1 to 7.3 during the second observation. To produce the best fit with COMPMAG model, we choose a decelrating velocity profile with index $\eta=0.5$ and terminal velocity $\beta_{0}=0.05$. In addition, we fixed the radius of the accretion column $r_{0}=0.25$ and albedo at NS surface A=1. This is the same assumption made by \cite{2012A&A...538A..67F} for the two other accretion-powered pulsars. The value of seed photon  electron temperature and optical depth with COMPMAG show similar trends between the two observations as with the COMPTT model.

\section{Time resolved spectral analysis} \label{sec:trsp}

To study the flaring activity seen in the observation of \igra, we extracted spectra from flaring ($\sim$1000~s) and off-flare intervals of the observation separately. Spectra were fitted in the 3.5--50.0~keV band using HIGHECUT+BBODY and COMPTT as continuum models. Resultant spectral parameters obtained from the fitting of these models are given in Table \ref{tab:cha}. The parameters of BBODY were poorly constrained in the flaring spectrum, so we fixed the blackbody temperature to the value found during off-flare duration. A lower value of column density $N_H=10\times10^{22}$ atoms cm$^{-2}$ was observed during flaring compared to  13.5$\times10^{22}$ atoms cm$^{-2}$ during off-flaring segments. A decrease in spectral index $\Gamma$, E$_{cut}$ and E$_{fold}$ is seen using empirical models. The effect of change in these spectral parameters indicates the hardening of spectra during flaring epochs. This becomes more evident with COMPTT model, where seed photon temperature and electron temperature are seen to decrease during flaring epochs along with the increase in the optical depth. An interesting interpretation can be extracted from the change in spectral parameters of the physical model in terms of comptonization parameter \(y\propto kT_{e}\tau^2\). The ratio $y_{flare}/y_{offflare}=1.8$ in this case implies that comptonization is more effective during the flare.

For \igrb, time resolved spectra from flaring and off-flaring segments were analyzed using the HIGHECUT and COMPTT models. The parameter values obtained for observation 1 and 2 from these intervals are given in Table \ref{tab:chbo} and \ref{tab:chbt} respectively. The column density $N_{H}$ value is consistent within errors for flaring and off-flaring spectra from both observations. During flaring in the first observation, the spectral index of the HIGHECUT model is smaller $1.10^{+0.09}_{-0.12}$ compared to $1.33^{+0.11}_{-0.11}$ during off-flare while $E_{cut}$ and $E_{fold}$ energies are consistent. The value of physical parameters, seed photon temperature and electron temperature are higher during flaring as obtained using the COMPTT model. More prominent spectral changes are observed during flaring in the second observation while the $E_{cut}$ energy decreases from $21^{+2}_{-4}$~keV to $8^{+0.6}_{-0.4}$~keV during flaring. Using COMPTT model, seed photon temperature was found to be $1.93^{+0.06}_{-0.06}$ keV while during off-flaring epochs this parameter could not be constrained. So, we fixed this parameter at the value obtained from the spectral data of full observation. The electron temperature was to be found higher during flaring compared to the off-flaring period. The value of optical depth does not change much during flaring. 

\section{Discussion} \label{sec:dis}

We present here a detailed timing and spectral analysis of two highly obscured HMXBs, \igra\ and \igrb\ ,  using archival \nus\ observations. \igra\ exhibits large variations in its orbital intensity profile without the presence of an eclipse in this source. \nus\ observations are taken around the minimum of this profile. Along with its well-known variable count rate, the \nus\ light curve shows a flare which is $\sim$1000~s long. We present energy-resolved pulse profile behavior and characterize its spectrum in the 3.5--50.0~keV band using different empirical and physical models. For \igrb, we present its orbital intensity profile using \swi\ data from 2005-2022. We present an out-of-eclipse hard X-ray spectrum for the first time for this source. Two very strong flares are seen in two \nus\ observations of \igrb. We present a time-resolved spectral analysis between the flaring and off-flaring epochs. 

\subsection{Flaring activity in \igra\ and \igrb}

The count rate during the \nus\ observation of \igra\ is moderate and highly variable. A flare is visible after $\sim$20 ks from the start of observation for $\sim$1000~s. The average luminosity of source during off-flare is $4\times10^{35}$ erg s$^{-1}$ (for a source distance of 3.5~kpc). The luminosity increased to 1.6$\times10^{36}$ erg s$^{-1}$ during the flare. Another flare was seen in 2004 using \xmm\ and \integral\ \citep{2006MNRAS.366..274R}. The duration of that flare was longer ($\sim$5 ks) compared to the flare in the \nus\ observation. The orbital phase of the flare in \xmm\ observation was at $\phi=0.42$, just before the minimum of orbital profile (taking MJD 54702.82310 as $\phi=0$ and $P_{orb}=8.991\pm0.001$ \citealt{2023NewA...9801942V}). The \nus\ flare occurred just after the minimum in the orbital profile ($\phi=0.523\pm0.001$). 

It is surprising to see flares in both the \nus\ observations of \igrb\ because flares in SFXTs occur sporadically and are difficult to catch. The average luminosity of the source during off-flare is $\sim$2$\times10^{35}$~erg~s$^{-1}$ (taking the recent estimates from \citealt{2015ApJ...808..140C}). The average luminosity during the first flare increased by a factor of 10 ($2\times10^{36}$~erg~s$^{-1}$). The second flare is even brighter than the first flare with an average luminosity of $8\times10^{36}$ erg s$^{-1}$. The dynamic range of source for the first and second flare are $\sim$35  and $\sim$135, respectively. Out of 25 reported flares (19 till 2006 \citep{2008A&A...487..619S}, 1 in 2008, 1 in 2009 \citep{2009A&A...502...21B}, 2 in 2012 \citep{2013MNRAS.429.2763S}, and 2 in this work), a large number of flares have dynamic range of $\sim$40 and only 4 have dynamic range $>$100. Hence, the first flare in \nus\ data is comparable to the general flaring activity in this source whereas the second flare is one of the very few bright flares in this source. We calculated the orbital phase of both flares using $P_{orb}=3.31961\pm0.00004$ days and epoch at 55081.571 MJD \citep{2015ApJ...808..140C}. The orbital phase of flare 1 ($T_{start}=58575.95$ MJD) is $\phi=0.346\pm0.002$. Hence, the first flare in the \nus\ data occurred at a location of known of flaring activity at 0.35 after the mid-eclipse or $\sim$0.65 prior to mid eclipse \citep{2009A&A...502...21B,2020ApJ...900...22S}. Recently, it was found that a giant flare had also occurred at this orbital phase in 2008 which was the brightest flare for this sources and second brightest among all the known SFXTs. It was proposed that flare at this location could be due to the presence of corotating interaction region (CIR) \citep{2020ApJ...900...22S}. This conjecture was based on the superorbital phase of giant flare and derived assuming a single arm CIR with an orbital period of 4.607 days. We checked the orbital phase and phase with respect to the CIR for flares in 2005, 2009, 2012, and 2019. While the orbital phases of these flares match completely, their phases with respect to CIR are different. We therefore conclude that other flares do not support CIR hypothesis. Instead, consistency in the orbital phase could be due to periastron passage. Future orbital parameter estimation for this source will be required to confirm this.

The second flare ($T_{start}=58574.81$ MJD) in the \nus\ data occurred at orbital phase $\phi=0.714\pm0.002$. This confirms that the flaring activity occurs at a site away from the site of first flare which was also indicated by \suz\ observation in 2012 \citep{2013MNRAS.429.2763S}. Similar to the \suz\ observation, the second flare is stronger in \nus\ data. However, its orbital phase is advanced by $\sim$0.15 compared to the previous observation. More observations near this orbital phase are required to check whether these flares are associated with a CIR or if they are coming from different mass clumps.

Flaring activity indicates the presence of inhomogenities or clumps in the wind of the companion star. We estimated the physical properties of these clumps responsible for flares in the \nus\ observations for both the sources. We considered these clumps to have roughly spherical shapes and duration of flares equal to characteristic time for NS to accrete this clump. Therefore, the clump size should be

\begin{equation}
    R_{cl}\sim\frac{T_{flare} v_{w}}{2}  
\end{equation}

\noindent Here $v_{w}$ is the wind velocity relative to NS. For O-B supergiant companions, typical value of $v_{w}$ is in range of 500-2000 km s$^{-1}$ \citep{2008A&ARv..16..209P}.\\

\noindent Then the mass density of clump can be estimated as:

\begin{equation}
    n_{cl}\sim\frac{N_H}{R_{cl}}    
\end{equation}

\noindent This implies the total mass accreted during flare is:

\begin{equation}
    M_{cl}= n_{cl}V_{cl} \sim n_{cl}R_{cl}^3
\end{equation}

\noindent Here $V_{cl}$ is the volume of clump. For \igra, we take \(v_{w}= 1000\) km s$^{-1}$ and \(N_{H} = 2\times10^{23}\) atoms cm$^{-2}$. This gives \(R_{cl}=5\times10^{10}\) cm and \(M_{cl}=9\times10^{21}\) g for this source. In case of \igrb, we assume same wind velocity value as in previous case and take column density, $N_{H}=9\times10^{22}$ atoms cm$^{-2}$ ($13\times10^{22}$ atoms cm$^{-2}$) for obs I  (Obs II) as obtained from the fitting above. This gives size of clump \(R_{cl1}\) and \(R_{cl2}\) to be $10^{11}$ cm. The mass of accreted clumps are $\sim$2$\times10^{22}$ g. 

The estimates for clump masses obtained here for both sources are $\sim$10000 times more massive than those obtained in hydrodynamic simulations of stellar winds from isolated supergiant stars \citep{2018A&A...611A..17S}. Additionally, the total energy released from accretion, ($L_{acc}\simeq0.1Mc^{2}$) of such clumps comes out to be $\sim$10$^{42}$~erg~s$^{-1}$. Considering an average flare luminosity in the range 10$^{36}$-10$^{37}$~erg~s$^{-1}$ and flare duration between 1000--3000~s, the energy released works out to be $\sim$10$^{39}$~erg~s$^{-1}$. Hence, a simplistic clumpy wind accretion model seems to be inconsistent with observations.

An alternate mechanism has been proposed for flaring activity in the quasi-spherical accretion regime at low accretion rates $\simeq$~$4\times10^{16}$~g~s$^{-1}$ \citep{2014MNRAS.442.2325S}. In this regime, the material between the magnetospheric radius and Bondi radius forms a quasi-static shell. Flares occurring at 1000~s timescale are caused due to the collapse of this shell by reconnection of large-scale magnetic field. The mass of this shell can be expressed as 

\begin{equation}
    \bigtriangleup M \approx 8\times10^{17}~L_{34}^{7/9}~v_{8}^{-3}~\mu_{30}^{-2/27} \rm g
\end{equation}

\noindent Here $L_{34}=L_{X}/10^{34}$~erg~s$^{-1}$ is the steady state luminosity of source, v$_{8}=v_{w}/10^{8}$ is normalized stellar wind velocity and $\mu_{30}=\mu/10^{30}$~G~cm$^{2}$ is the magnetic moment of NS (very weak dependence on this parameter). Using $v_{8}=1$, $L_{34}=40$ for \igra ~($L_{34}=20$ for \igrb), and $\mu_{30}=1$, we obtain the mass of the shell to be $1.4\times$10$^{19}$~g~($0.8\times10^{19}$~g for \igrb). If the entire shell is completely accreted during this event then the total energy released comes out to be $\sim$10$^{39}$~erg~s$^{-1}$ which is agreement with observational estimates.

\subsection{Behavior of pulsation in \igra}

The pulse profile of \igra\ obtained from \nus\ observation contains a single broad peak along with an interpulse consistent with the pulse profile observed in soft X-rays using \xmm. The pulsed fraction in the 3.0--79.0~keV band is $\sim$12\% which is smaller compared to $\sim$18\% in soft X-rays using \xmm\ (0.2-10~keV) observations. This indicates that the pulsation are less prominent in hard X-ray band. This behavior is confirmed in energy resolved pulse profiles with a decreasing pulsed fraction from 22\% in the 3-7~keV band to 10\% in 20-30~keV band.  In contrast, the \integral\ observation in 2004 \citep{2003A&A...407L..41R} had reported a constant pulse fraction upto 20~keV. However, the data point corresponding to the 10-20 keV band in their results is shown with large uncertainty. The low pulsed fraction of the pulse profile is compatible with either a small misalignment between the rotation and magnetic axis or low viewing angle with the equator neutron star. Further, the energy dependence suggest that the hard X-ray regions are more compact. 

\subsection{Hard X-ray spectral properties of \igra}

The continuum parameters ($\Gamma$, $E_{cut}$, and $E_{fold}$) of the HIGHECUT model in \igra\ are similar to those observed in other supergiant HMXBs like \hma\ \citep{2014ApJ...792...14H},\hmb\ \citep{2019ApJ...880...61V}, and  \hmc\ \citep{2022A&A...660A..19D}. The seed photon temperature and electron temperature values are consistent with values reported using joint \xmm\ and \integral\ study \citep{2006MNRAS.366..274R}. However, we have obtained a higher optical depth ($\tau$) compared to their study. Iron emission line is detected with low equivalent with of 75 eV. Such a weak iron line at orbital phase near the minimum is consistent with orbital phase variation of spectral parameters of this source \citep{2009AIPC.1126..313H}. No cyclotron line is detected in hard X-ray spectra either with \nus\ data (3.5--50.0~keV) or \xmm+\integral\ (0.6-80.0 keV). This implies that either the magnetic field strength of \igra\ is such that the expected CRSF feature lies outside the energy band covered by these instruments or a weak CRSF is formed which is undetected. We now use some indirect methods to estimate the magnetic field of this source using its observed properties.

For an accreting pulsar with a long spin period such as \igra, magnetic field strength can be calculated in the fast rotator regime of standard accretion disk model \citep{1979ApJ...234..296G}. Presence of a persistent or temporary disk has been inferred in some HMXB sources like \sm\ \citep{1977ApJ...217..186H}, \lm\ \citep{1989A&A...223..154H}, \gx\ \citep{2019A&A...629A.101N}, and \oao\ \citep{2019IAUS..346..178S}.  The accreted matter can exert torque on the pulsar to either increase or decrease its spin, and a spin equilibrium state is arrived after multiple spin-up or spin-down episodes \citep{2014MNRAS.437.3664H}. For fastness parameter $\omega_{s}=0.35$ \citep{1979ApJ...234..296G} the equilibrium spin is given as:

\begin{equation}
   P_{eq}=67s\zeta^{3/2}M_{1.4}^{-2/3}R_{6}^{-2/7}L_{37}^{-3/7}\Bigl(\frac{B}{10^{13}G}\Bigr)^{6/7}
\end{equation}

\noindent Where $\zeta=1$ for magnetically dominated matter, $M_{1.4}=M_{NS}/1.4M_{\odot}$, $L_{37}=L_{X}/10^{37}$ erg s$^{-1}$ and $R_{6}=R_{NS}/10^{6}$ cm. Considering the off-flare luminosity of source is $2\times10^{35}$ erg $s^{-1}$ and P$_{eq}=1306$ s, the magnetic field strength of the NS in \igra\ is $\sim6\times10^{13}$ G. For higher value of fastness parameter (for eg. $\omega_{s}=1$), the inferred magnetic field strength increases to $2\times10^{14}$ G. 

Another estimate of magnetic field can be obtained using quasi-spherical settling model \citep{2012MNRAS.420..216S,2012int..workE..22P}. This model is applicable for HMXB sources like \igra\ hosting a slowly rotating NS and moderate luminosity. In the quasi-spherical accretion regime, matter does not cool down rapidly at the accretion radius and settles slowly on to the magnetosphere in the form of a quasi-static shell. The plasma in this shell has strong convective motions and follow an iso-angular-momentum rotation distribution ($n\approx2$ case). The spin period change of NS is determined by net torques from specific angular momentum of wind material and total angular momentum of plasma motion in the shell. The NS is found to rotate near equilibrium periods in many sources. For such sources, the equilibrium spin period $P_{eq}^{\ast}$ is related to orbital period P$_{orb}$, the mass accretion rate $\dot{M}_{16}=M/10^{16}$~g~s$^{-1}$, NS magnetic field B (through $\mu_{30}$) and relative stellar wind velocity $v_{8}=v_{w}/(1000 km/s)$ through the following equation:

\begin{equation}
    P_{eq}^{\ast}=1300~\mu_{30}^{12/11}~(P_{orb}/10d)~\dot{M}_{16}^{-4/11}~v_{8}^{4}
\end{equation} 

\noindent Using $P_{eq}^{\ast}=1306$~s, $P_{orb}=8.991$ days, M$_{16}=0.4$, v$_{8}=1$, we get $B=8\times10^{11}$ G. If \igra\ is indeed accreting in the quasi-spherical accretion regime (as discussed above for the flaring activity observed using \nus\ ), the CRSF feature should be formed at $\sim$7 keV. Non-detection of such a feature in the \nus\ spectrum could indicate that the fundamental CRSF might be very shallow and filled by spawned photons from higher harmonics \citep{2007A&A...472..353S}. Such a possibility has been discussed for \hmf\ \citep{2012A&A...538A..67F} and \gs\ \citep{2013ApJ...775...65M}  due to non-detection of CRSF.

\subsection{Hard X-ray spectrum of \igrb}

Results obtained from fitting the out-of-eclipse \nus\ spectrum of \igrb\ provides, for the first time, some insight into the origin of direct emission from NS in hard X-rays (10--70.0~keV). Thermal comptonization models provide good fits to the continuum spectrum and lends physical interpretation. The seed photon with temperature $T_{0}=1.51\pm0.08$~keV which originate near the surface of NS are compton up-scattered by hotter plasma electrons with temperature kT$=10.5^{+0.6}_{-0.5}$ at higher energies. The broadband spectrum in 3.5-70.0 keV takes a hard powerlaw shape ($\Gamma=1.15\pm0.05$) with an exponential cutoff at 6.6$^{+0.6}_{-0.5}$~keV. The spectral parameters of this source are similar to other supergiant HMXB line \igrc\ \citep{2019ApJ...879...34C}, \igrd\ \citep{2016ApJ...823..146B}, and \igre\ \citep{2015MNRAS.447.2274B}.

No CSRF is detected in the 3.5-70.0 keV broadband spectrum of this source. If the CRSF feature lies outside this energy range then future broadband observations  using HEX-P (0.1-150.0~keV) \citep{2019BAAS...51g.166M}  will prove to be useful in order to probe and constrain the magnetic field strength in this source. Alternatively, the CRSF in this source might be shallower than the CRSF in \igre\ \citep{2015MNRAS.447.2274B} and \igrc\ \citep{2016ApJ...823..146B}, the only other SFXT sources where the presence of CRSF has been indicated. Understanding the inherent shallow nature of CRSF in SFXT sources can serve as an interesting direction for future theoretical studies in the field of pulsar accretion.

\section*{Acknowledgement} This research work has made use of \nus\ Data Analysis Software (NuSTARDAS) which has been jointly developed by the ASI Science Data Center (ASDC, Italy) and the Calfornia Institute of Technology, USA.  We thank Prof. Biswajit Paul for giving useful comments and suggestions on the draft of this paper. We also thank Prof. Konstantin A. Postnov for help with understanding of the Quasi-spherical accretion model. 

\section*{Data Availability} All the data used in this work is publicly available. \nus\ data can be downloaded from HEASARC archive (\url{https://heasarc.gsfc.nasa.gov/cgi-bin/W3Browse/w3browse.pl}) and the \swi\ light curves of both these sources can be downloaded from the database of hard X-ray transients (\url{https://swift.gsfc.nasa.gov/results/transients/}).

\begin{table*}
    \centering
    \caption{Best fit spectral parameters of \igra\ spectrum for several phenomenological models.}
    \label{tab:pha}
    \begin{tabular}{|p{2cm}||p{1.5cm}|p{1.5cm}|p{1.5cm}|p{1.5cm}|p{1.5cm}|p{1.5cm}|p{1.5cm}}
     \hhline{========}
     Parameter & \multicolumn{7}{c}{Continuum spectral model}  \\
    \hline
              & \multicolumn{2}{c}{Highecut}                        & \multicolumn{2}{c}{FDCUT}                         & \multicolumn{2}{c}{NEWHCUT}                   & Cutoffpl            \\
    \hline     
              &  Without BBody           & With BBody                & Without BBody             & With BBody            & Without BBody         & With BBody            & Without BBody       \\
     \hline
     N$_H^a$      &  $30^{+1}_{-1}$             & $21^{+1}_{-2}$            & $22^{+1}_{-1}$            &$19^{+2}_{-2}$         & $29^{+1}_{-1}$        & $21^{+2}_{-2}$        & $17.6^{+0.9}_{-1.0}$       \\
     \hline
     $\Gamma$   &  $1.32^{+0.02}_{-0.02}$  & $1.14^{+0.05}_{-0.05}$    & $0.83^{+0.08}_{-0.09}$    & $0.7^{+0.1}_{-0.2}$   & $1.29^{+0.03}_{-0.03}$& $1.09^{+0.06}_{-0.06}$ & $0.35^{+0.04}_{-0.04}$\\
     E$_{cut}$ (keV)  &  $21.7^{+0.6}_{-1.0}$    & $25.6^{+0.6}_{-1.6}$      & $17^{+4}_{-5}$            & $23^{+4}_{-5}$        & $21^{+1}_{-1}$        & $24.8^{+0.9}_{-0.8}$  & $14.0^{+0.5}_{-0.4}$ \\
     E$_{fold}$ (keV) &  $21.0^{+1.1}_{-0.9}$     & $17^{+1}_{-1}$            & $12.4^{+0.7}_{-0.7}$      & $10.3^{+0.8}_{-0.8}$  & $20^{+1}_{-1}$        &$17^{+1}_{-1}$         &  \\
     \hline
     kT$_{BB}$ (keV) &                          & $3.4^{+0.2}_{-0.3}$       &                           &  $2.3^{+0.3}_{-0.3}$  &                       & $3.2^{+0.2}_{-0.2}$   &         \\
     \hline
     E$_{Fe}$ (keV)        &  $6.38^{+0.07}_{-0.06}$  & $6.36^{+0.06}_{-0.05}$  & $6.39^{+0.03}_{-0.02}$ & $6.36^{+0.04}_{-0.05}$& $6.39^{+0.04}_{-0.05}$ & $6.36^{+0.04}_{-0.04}$& $6.37^{+0.01}_{-0.04}$ \\
     EWQ (eV)                   & 30$^{+10}_{-10}$      &   76$^{+13}_{-15}$ & $71^{+14}_{-12}$     & 76$^{+16}_{-11}$ & 37$^{+11}_{-13}$ & 76$^{+14}_{-14}$ & 88$^{+13}_{-14}$ \\
     \hline
     F$^{b}_{3.5-50.0 keV}$   & 2.57$^{+0.02}_{-0.02}$ & 2.55$^{+0.02}_{-0.04}$& 2.55$^{+0.02}_{-0.03}$ & 2.553$^{+0.005}_{-0.356}$ & 2.56$^{+0.02}_{-0.02}$ & 2.56$^{+0.01}_{-0.05}$ & 2.55$^{+0.02}_{-0.03}$ \\ 
     \hline
     $\chi^2$ (dof)   &   1.39 (1188)            & 1.15 (1186)               &  1.20 (1188)             &   1.17 (1186)         & 1.37 (1188)                  &  1.15 (1186)                 & 1.23 (1189)\\ 
     \hline
    \end{tabular}\\
    \begin{flushleft}
    $^a$ Hydrogen column density is measured in units of 10$^{22}$ atoms $cm^{-2}$. $^{b}$ Flux of \igra\ in 3.5-50.0 keV band is in units of $10^{-10}$ erg cm$^{-2}$.
    \end{flushleft}
\end{table*}

\begin{table}
    \centering
    \caption{Best fit spectral parameters of \igra\ spectrum for three physical models. In case of COMPMAG model the fixed parameters are $\eta=0.5$, $\beta_{0}=0.05$, $r_{0}=0.25$, and $A=1$.}
    \label{tab:tha}
    \begin{tabular}{|p{1.8cm}|p{1.3cm}|p{1.3cm}|p{1.3cm}|}
    \hhline{====}
    Parameter           &   COMPTT                  & NTHcomp              & COMPMAG                    \\
    \hline
      N$_H^a$             & $15^{+1}_{-1}$            & $17^{+1}_{-1}$        & $16.9^{+0.9}_{-0.8}$     \\
      \hline
      T$_0$ (keV)       & $1.84^{+0.07}_{-0.07}$    &                       &                            \\
      kT    (keV)       & $9.4^{+0.1}_{-0.1}$       &                       &                            \\
      $\tau$            & $10.8^{+0.2}_{-0.2}$      &                       &  $0.666^{+0.011}_{-0.006}$ \\
      $\Gamma$          &                           & $1.44^{+0.01}_{-0.01}$&                            \\
      kT$_{e}$ (keV)    &                           & $10.0^{+0.2}_{-0.2}$  & $5.98^{+0.05}_{-0.10}$     \\
      kT$_{BB}$ (keV)   &                           & $1.8^{+0.1}_{-0.1}$   & $1.56^{+0.07}_{-0.06}$     \\
      \hline
      E$_{Fe}$ (keV)     & $6.381^{+0.008}_{-0.013}$ & $6.38^{+0.03}_{-0.02}$&  $6.383^{+0.006}_{-0.013}$ \\
      EQW (keV)         & 78$^{+12}_{-14}$           & 78$^{+14}_{-13}$        & 73$^{+11}_{-14}$        \\
      \hline
      F$^{b}_{3.5-50.0 keV}$ & 2.55$^{+0.02}_{-0.02}$& 2.56$^{+0.01}_{-0.04}$ & 2.53$^{+0.07}_{-0.02}$ \\
      \hline
      $\chi^{2}$ (dof)       &  1.19 (1188)                     &   1.18 (1188)               &   1.19 (1188)            \\  
      \hline
    \end{tabular}
    \begin{flushleft}
    $^a$ Hydrogen column density is measured in units of 10$^{22}$ atoms $cm^{-2}$. $^{b}$ Flux of \igra\ in 3.5-50.0 keV band is in units of $10^{-10}$ erg cm$^{-2}$.
    \end{flushleft}    
\end{table}

\begin{table}
    \centering
    \caption{Best-fit spectral parameters of \igra\ during time-resolved periods.}
    \label{tab:cha}
    \begin{tabular}{|p{1.8cm}|p{1.3cm}|p{1.3cm}|p{1.3cm}|p{1.3cm}|}
    \hhline{=====}
    Paramater           & \multicolumn{2}{c}{HIGHECUT}                      & \multicolumn{2}{c}{COMPTT}                        \\
    \hline
                        & Flare                 & Off-Flare                 &   Flare               & Off-Flare                 \\
    \hline
    N$_{H}^{a}$         & $10^{+4}_{-3}$        & $13.5^{+0.7}_{-1.4}$      & $8^{+4}_{-4}$         & $15^{+2}_{-1}$            \\
    \hline
    $\Gamma$            & $0.7^{+0.2}_{-0.2}$   & $1.11^{+0.04}_{-0.09}$    &                       &                           \\
    $E_{cut}$ (keV)     & $21^{+1}_{-1}$        & $25.8^{+0.6}_{-1.0}$      &                       &                           \\
    $E_{fold}$ (keV)    & $13^{+2}_{-1}$        & $16.4^{+0.8}_{-1.1}$      &                       &                           \\
    \hline
    kT$_{BB}$   (keV)   & 3.2                   & $3.2^{+0.1}_{-0.2}$       &                       &                           \\
    \hline
    T$_{0}$             &                       &                           & $1.3^{+0.3}_{-0.4}$   & $1.65^{+0.07}_{-0.06}$    \\
    kT (keV)            &                       &                           & $7.2^{+0.3}_{-0.3}$   & $8.5^{+0.1}_{-0.1}$       \\
    $\tau$              &                       &                           & $16^{+1}_{-1}$        & $11.1^{+0.2}_{-0.2}$      \\
    \hline
    F$^{b}_{3.5-50.0 keV}$  & 10.1$^{+0.1}_{-0.4}$ & 2.41$^{+0.01}_{-0.05}$     & 10.0$^{+0.2}_{-0.4}$   & 2.41$^{+0.01}_{-0.02}$    \\
    \hline
    $\chi^{2}$ (dof)    & 1.13 (249)            &   1.16 (1171)              & 1.12 (249)             & 1.21 (1171)              \\
    \hline
    \end{tabular}\\
    \begin{flushleft}
    $^a$ Hydrogen column density is measured in units of 10$^{22}$ atoms $cm^{-2}$. $^{b}$ Flux of \igra\ in 3.5-50.0 keV band is in units of $10^{-10}$ erg cm$^{-2}$.
    \end{flushleft}
\end{table}

\begin{table*}
    \centering
    \caption{Best fit spectral parameters of \igrb\ spectrum during the \nus\ observations.}
    \label{tab:phbo}
    \begin{tabular}{|p{1.6cm}|p{1.2cm}|p{1.2cm}|p{1.2cm}|p{1.2cm}|p{1.2cm}|p{1.3cm}}
    \hhline{=======}
    Parameter           & \multicolumn{2}{c}{HIGHECUT}                          & \multicolumn{2}{c}{COMPTT}                   & \multicolumn{2}{c}{COMPMAG}\\
    \hline
                        & Obs 1                     & Obs 2                     & Obs 1                & Obs 2                 & Obs 1          & Obs 2         \\                            \\
    \hline                    
    N$_H^a$             &   $9^{+2}_{-2}$           & $13^{+1}_{-1}$            &   $3^{+2}_{-2}$      & $5^{+1}_{-1}$         & $4.38^{+0.97}_{-0.02}$ &  $6.2^{+1.1}_{-0.8}$   \\
    \hline
    $\Gamma$            &   $1.16^{+0.08}_{-0.08}$  & $1.36^{+0.05}_{-0.08}$    &                      &                       &    &   \\
    E$_{cut}$ (keV)     &   $6.6^{+0.6}_{-0.5}$     & $7.7^{+0.5}_{-0.9}$       &                      &                       &    &   \\
    E$_{fold}$ (keV)    &   $27^{+4}_{-3}$          & $30^{+3}_{-3}$            &                      &                       &    &   \\
    \hline
    T$_0$  (keV)        &                           &                           &$1.51^{+0.08}_{-0.08}$& $1.55^{+0.05}_{-0.05}$ & $1.46^{+0.04}_{-0.04}$  & $1.49^{+0.06}_{-0.06}$  \\
    kT      (keV)       &                           &                           & $10.5^{+0.6}_{-0.5}$ & $10.9^{+0.5}_{-0.4}$   & $7.9^{+0.2}_{-0.2}$     & $8.4^{+0.4}_{-0.4}$      \\
    $\tau$              &                           &                           & $8.0^{+0.3}_{-0.3}$  & $7.3^{+0.2}_{-0.3}$    & $0.43^{+0.02}_{-0.02}$   &     $0.38^{+0.01}_{-0.01}$  \\
    \hline
    F$^{b}_{3.5-70.0 keV}$  & 1.76$^{+0.02}_{-0.10}$ & 2.60$^{+0.03}_{-0.07}$  & 1.72$^{+0.02}_{-0.05}$ & 2.55$^{+0.03}_{-0.04}$ & 1.75$^{+0.01}_{-0.44}$ & 2.59$^{+0.74}_{-0.03}$ \\

    \hline
    $\chi^2$ (dof)           &  1.06 (721)            &  1.17 (947)                     & 1.09 (721)               & 1.22 (947)                   & 1.08 (721)                 &1.20 (947) \\
    \hline
    \end{tabular}\\
    \begin{flushleft}
    $^a$ Hydrogen column density is measured in units of 10$^{22}$ atoms $cm^{-2}$. $^{b}$ Flux of \igrb\ in 3.5-70.0 keV band is in units of $10^{-10}$ erg cm$^{-2}$.
    \end{flushleft}
\end{table*}

\begin{table}
    \centering
    \caption{Best-fit spectral parameters of \igrb\ during time-resolved segments in the first observation.}
    \label{tab:chbo}
    \begin{tabular}{|p{1.8cm}|p{1.3cm}|p{1.3cm}|p{1.3cm}|p{1.3cm}|}
    \hhline{=====}
    Parameter       & \multicolumn{2}{c}{HIGHECUT}                  & \multicolumn{2}{c}{COMPTT}                    \\
    \hline
                    & Flare                 & Off-Flare             &  Flare                &   Off-Flare           \\
    \hline
    N$_H^a$         & $12^{+2}_{-2}$        & $8^{+2}_{-2}$         & $4^{+2}_{-2}$         & $4^{+3}_{-3}$         \\
    \hline
    $\Gamma$        & $1.10^{+0.09}_{-0.12}$& $1.33^{+0.11}_{-0.11}$&                       &                       \\
    E$_{cut}$ (keV) & $7.8^{+0.7}_{-1.1}$   & $6.4^{+0.7}_{-1.2}$   &                       &                       \\
    E$_{fold}$ (keV)& $29^{+5}_{-4}$        & $26^{+6}_{-4}$        &                       &                       \\
    \hline
    T$_{0}$ (keV)   &                       &                       & $2.0^{+0.1}_{-0.1}$   & $1.4^{+0.1}_{-0.2}$   \\
    kT (keV)        &                       &                       & $12.9^{+1.0}_{-0.8}$  & $10.8^{+1.0}_{-0.8}$   \\
    $\tau$          &                       &                       & $7.6^{+0.4}_{-0.5}$   & $7.4^{+0.5}_{-0.5}$   \\
    \hline
    F$^{b}_{3.5-60.0 keV}$ & 9.4$^{+0.1}_{-0.9}$& 0.81$^{+0.01}_{-0.10}$ & 9.3$^{+0.2}_{-0.2}$& 0.79$^{+0.01}_{-0.03}$ \\
    \hline
    $\chi^2$ (dof)   & 1.02 (532)            & 1.04 (497)           & 1.06 (532)             & 1.03 (497)                 \\
    \hline
    \end{tabular}
    \begin{flushleft}
    $^a$ Hydrogen column density is measured in units of 10$^{22}$ atoms $cm^{-2}$. $^{b}$ Flux of \igrb\ in 3.5-60.0 keV band is in units of $10^{-10}$ erg cm$^{-2}$. 
    \end{flushleft}
\end{table}

\begin{table}
    \centering
    \caption{Best-fit spectral parameters of \igrb\ during time-resolved segments in the second observation.}
    \label{tab:chbt}
    \begin{tabular}{|p{1.8cm}|p{1.3cm}|p{1.3cm}|p{1.3cm}|p{1.3cm}|}
    \hhline{=====}
    Parameter       & \multicolumn{2}{c}{HIGHECUT}                      & \multicolumn{2}{c}{COMPTT}                        \\
    \hline
                    & Flare                 & Off-Flare                 &  Flare                &   Off-Flare               \\
    \hline
    N$_{H}^a$         & $14^{+1}_{-1}$        & $17^{+1}_{-1}$          & $3^{+1}_{-1}$         & $8^{+1}_{-1}$            \\
    \hline
    $\Gamma$        & $1.35^{+0.06}_{-0.05}$& $1.74^{+0.04}_{-0.04}$    &                       &                           \\
    E$_{cut}$ (keV) & $8.1^{+0.6}_{-0.4}$   & $21^{+2}_{-2}$                       &                       &                           \\
    E$_{fold}$ (keV)& $30^{+3}_{-2}$        & $33^{+9}_{-6}$            &                       &                           \\
    \hline
    T$_{0}$ (keV)   &                       &                           & $1.93^{+0.06}_{-0.06}$& 1.4                      \\
    kT (keV)        &                       &                           & $12.7^{+0.8}_{-0.7}$  & $10.9^{+0.7}_{-0.6}$       \\
    $\tau$          &                       &                           & $6.4^{+0.3}_{-0.3}$   & $7.6^{+0.4}_{-0.4}$       \\
    \hline
    F$^{b}_{3.5-60.0 keV}$ & 30.5$^{+0.3}_{-0.9}$& 0.78$^{+0.02}_{-0.03}$& 30.6$^{+0.2}_{-0.4}$ & 0.76$^{+0.02}_{-0.02}$          \\
    \hline
    $\chi^2$ (dof)       & 1.20 (826)            & 1.08 (600)                 & 1.24 (826)                  & 1.06 (600)              \\
    \hline
    \end{tabular}\\
    \begin{flushleft}
    $^a$ Hydrogen column density is measured in units of 10$^{22}$ atoms $cm^{-2}$. $^{b}$ Flux of \igrb\ in 3.5-60.0 keV band is in units of $10^{-10}$ erg cm$^{-2}$. 
    \end{flushleft}
\end{table}

\bibliographystyle{mnras}
\bibliography{newref}

\begin{thebibliography}{}
\makeatletter
\relax
\def\mn@urlcharsother{\let\do\@makeother \do\$\do\&\do\#\do\^\do\_\do\%\do\~}
\def\mn@doi{\begingroup\mn@urlcharsother \@ifnextchar [ {\mn@doi@}
  {\mn@doi@[]}}
\def\mn@doi@[#1]#2{\def\@tempa{#1}\ifx\@tempa\@empty \href
  {http://dx.doi.org/#2} {doi:#2}\else \href {http://dx.doi.org/#2} {#1}\fi
  \endgroup}
\def\mn@eprint#1#2{\mn@eprint@#1:#2::\@nil}
\def\mn@eprint@arXiv#1{\href {http://arxiv.org/abs/#1} {{\tt arXiv:#1}}}
\def\mn@eprint@dblp#1{\href {http://dblp.uni-trier.de/rec/bibtex/#1.xml}
  {dblp:#1}}
\def\mn@eprint@#1:#2:#3:#4\@nil{\def\@tempa {#1}\def\@tempb {#2}\def\@tempc
  {#3}\ifx \@tempc \@empty \let \@tempc \@tempb \let \@tempb \@tempa \fi \ifx
  \@tempb \@empty \def\@tempb {arXiv}\fi \@ifundefined
  {mn@eprint@\@tempb}{\@tempb:\@tempc}{\expandafter \expandafter \csname
  mn@eprint@\@tempb\endcsname \expandafter{\@tempc}}}

\bibitem[\protect\citeauthoryear{{Aftab}, {Paul}  \& {Kretschmar}}{{Aftab}
  et~al.}{2019}]{2019ApJS..243...29A}
{Aftab} N.,  {Paul} B.,   {Kretschmar} P.,  2019, \mn@doi [\apjs]
  {10.3847/1538-4365/ab2a77}, \href
  {https://ui.adsabs.harvard.edu/abs/2019ApJS..243...29A} {243, 29}

\bibitem[\protect\citeauthoryear{{Becker} \& {Wolff}}{{Becker} \&
  {Wolff}}{2007}]{2007ApJ...654..435B}
{Becker} P.~A.,  {Wolff} M.~T.,  2007, \mn@doi [\apj] {10.1086/509108}, \href
  {https://ui.adsabs.harvard.edu/abs/2007ApJ...654..435B} {654, 435}

\bibitem[\protect\citeauthoryear{{Bhalerao} et~al.,}{{Bhalerao}
  et~al.}{2015}]{2015MNRAS.447.2274B}
{Bhalerao} V.,  et~al., 2015, \mn@doi [\mnras] {10.1093/mnras/stu2495}, \href
  {https://ui.adsabs.harvard.edu/abs/2015MNRAS.447.2274B} {447, 2274}

\bibitem[\protect\citeauthoryear{{Bodaghee} et~al.,}{{Bodaghee}
  et~al.}{2016}]{2016ApJ...823..146B}
{Bodaghee} A.,  et~al., 2016, \mn@doi [\apj] {10.3847/0004-637X/823/2/146},
  \href {https://ui.adsabs.harvard.edu/abs/2016ApJ...823..146B} {823, 146}

\bibitem[\protect\citeauthoryear{{Boldin}, {Tsygankov}  \&
  {Lutovinov}}{{Boldin} et~al.}{2013}]{2013AstL...39..375B}
{Boldin} P.~A.,  {Tsygankov} S.~S.,   {Lutovinov} A.~A.,  2013, \mn@doi
  [Astronomy Letters] {10.1134/S1063773713060029}, \href
  {https://ui.adsabs.harvard.edu/abs/2013AstL...39..375B} {39, 375}

\bibitem[\protect\citeauthoryear{{Bozzo}, {Falanga}  \& {Stella}}{{Bozzo}
  et~al.}{2008}]{2008ApJ...683.1031B}
{Bozzo} E.,  {Falanga} M.,   {Stella} L.,  2008, \mn@doi [\apj]
  {10.1086/589990}, \href
  {https://ui.adsabs.harvard.edu/abs/2008ApJ...683.1031B} {683, 1031}

\bibitem[\protect\citeauthoryear{{Bozzo}, {Giunta}, {Stella}, {Falanga},
  {Israel}  \& {Campana}}{{Bozzo} et~al.}{2009}]{2009A&A...502...21B}
{Bozzo} E.,  {Giunta} A.,  {Stella} L.,  {Falanga} M.,  {Israel} G.,
  {Campana} S.,  2009, \mn@doi [\aap] {10.1051/0004-6361/200912131}, \href
  {https://ui.adsabs.harvard.edu/abs/2009A&A...502...21B} {502, 21}

\bibitem[\protect\citeauthoryear{{Bozzo}, {Oskinova}, {Feldmeier}  \&
  {Falanga}}{{Bozzo} et~al.}{2016}]{2016A&A...589A.102B}
{Bozzo} E.,  {Oskinova} L.,  {Feldmeier} A.,   {Falanga} M.,  2016, \mn@doi
  [\aap] {10.1051/0004-6361/201628341}, \href
  {https://ui.adsabs.harvard.edu/abs/2016A&A...589A.102B} {589, A102}

\bibitem[\protect\citeauthoryear{{Bozzo}, {Bernardini}, {Ferrigno}, {Falanga},
  {Romano}  \& {Oskinova}}{{Bozzo} et~al.}{2017}]{2017A&A...608A.128B}
{Bozzo} E.,  {Bernardini} F.,  {Ferrigno} C.,  {Falanga} M.,  {Romano} P.,
  {Oskinova} L.,  2017, \mn@doi [\aap] {10.1051/0004-6361/201730398}, \href
  {https://ui.adsabs.harvard.edu/abs/2017A&A...608A.128B} {608, A128}

\bibitem[\protect\citeauthoryear{{Burderi}, {Di Salvo}, {Robba}, {La Barbera}
  \& {Guainazzi}}{{Burderi} et~al.}{2000}]{burderi2000}
{Burderi} L.,  {Di Salvo} T.,  {Robba} N.~R.,  {La Barbera} A.,   {Guainazzi}
  M.,  2000, \mn@doi [\apj] {10.1086/308336}, \href
  {http://adsabs.harvard.edu/abs/2000ApJ...530..429B} {530, 429}

\bibitem[\protect\citeauthoryear{{Chaty}, {Rahoui}, {Foellmi}, {Tomsick},
  {Rodriguez}  \& {Walter}}{{Chaty} et~al.}{2008}]{2008A&A...484..783C}
{Chaty} S.,  {Rahoui} F.,  {Foellmi} C.,  {Tomsick} J.~A.,  {Rodriguez} J.,
  {Walter} R.,  2008, \mn@doi [\aap] {10.1051/0004-6361:20078768}, \href
  {https://ui.adsabs.harvard.edu/abs/2008A&A...484..783C} {484, 783}

\bibitem[\protect\citeauthoryear{{Coleiro}, {Chaty}, {Zurita Heras}, {Rahoui}
  \& {Tomsick}}{{Coleiro} et~al.}{2013}]{2013A&A...560A.108C}
{Coleiro} A.,  {Chaty} S.,  {Zurita Heras} J.~A.,  {Rahoui} F.,   {Tomsick}
  J.~A.,  2013, \mn@doi [\aap] {10.1051/0004-6361/201322382}, \href
  {https://ui.adsabs.harvard.edu/abs/2013A&A...560A.108C} {560, A108}

\bibitem[\protect\citeauthoryear{{Coley}, {Corbet}  \& {Krimm}}{{Coley}
  et~al.}{2015}]{2015ApJ...808..140C}
{Coley} J.~B.,  {Corbet} R. H.~D.,   {Krimm} H.~A.,  2015, \mn@doi [\apj]
  {10.1088/0004-637X/808/2/140}, \href
  {https://ui.adsabs.harvard.edu/abs/2015ApJ...808..140C} {808, 140}

\bibitem[\protect\citeauthoryear{{Coley}, {Corbet}, {F{\"u}rst}, {Huxtable},
  {Krimm}, {Pearlman}  \& {Pottschmidt}}{{Coley}
  et~al.}{2019}]{2019ApJ...879...34C}
{Coley} J.~B.,  {Corbet} R. H.~D.,  {F{\"u}rst} F.,  {Huxtable} G.,  {Krimm}
  H.~A.,  {Pearlman} A.~B.,   {Pottschmidt} K.,  2019, \mn@doi [\apj]
  {10.3847/1538-4357/ab223c}, \href
  {https://ui.adsabs.harvard.edu/abs/2019ApJ...879...34C} {879, 34}

\bibitem[\protect\citeauthoryear{{Corbet} \& {Krimm}}{{Corbet} \&
  {Krimm}}{2013}]{2013ApJ...778...45C}
{Corbet} R. H.~D.,  {Krimm} H.~A.,  2013, \mn@doi [\apj]
  {10.1088/0004-637X/778/1/45}, \href
  {https://ui.adsabs.harvard.edu/abs/2013ApJ...778...45C} {778, 45}

\bibitem[\protect\citeauthoryear{{Davidson} \& {Ostriker}}{{Davidson} \&
  {Ostriker}}{1973}]{1973ApJ...179..585D}
{Davidson} K.,  {Ostriker} J.~P.,  1973, \mn@doi [\apj] {10.1086/151897}, \href
  {https://ui.adsabs.harvard.edu/abs/1973ApJ...179..585D} {179, 585}

\bibitem[\protect\citeauthoryear{{Diez} et~al.,}{{Diez}
  et~al.}{2022}]{2022A&A...660A..19D}
{Diez} C.~M.,  et~al., 2022, \mn@doi [\aap] {10.1051/0004-6361/202141751},
  \href {https://ui.adsabs.harvard.edu/abs/2022A&A...660A..19D} {660, A19}

\bibitem[\protect\citeauthoryear{{Dorman} \& {Arnaud}}{{Dorman} \&
  {Arnaud}}{2001}]{2001ASPC..238..415D}
{Dorman} B.,  {Arnaud} K.~A.,  2001, in {Harnden} F.~R. J.,  {Primini} F.~A.,
  {Payne} H.~E.,  eds,  Astronomical Society of the Pacific Conference Series
  Vol. 238, Astronomical Data Analysis Software and Systems X. p.~415

\bibitem[\protect\citeauthoryear{{Ducci}, {Sidoli}, {Mereghetti}, {Paizis}  \&
  {Romano}}{{Ducci} et~al.}{2009}]{2009MNRAS.398.2152D}
{Ducci} L.,  {Sidoli} L.,  {Mereghetti} S.,  {Paizis} A.,   {Romano} P.,  2009,
  \mn@doi [\mnras] {10.1111/j.1365-2966.2009.15265.x}, \href
  {https://ui.adsabs.harvard.edu/abs/2009MNRAS.398.2152D} {398, 2152}

\bibitem[\protect\citeauthoryear{{Farinelli}, {Ceccobello}, {Romano}  \&
  {Titarchuk}}{{Farinelli} et~al.}{2012}]{2012A&A...538A..67F}
{Farinelli} R.,  {Ceccobello} C.,  {Romano} P.,   {Titarchuk} L.,  2012,
  \mn@doi [\aap] {10.1051/0004-6361/201118008}, \href
  {https://ui.adsabs.harvard.edu/abs/2012A&A...538A..67F} {538, A67}

\bibitem[\protect\citeauthoryear{{Garc{\'\i}a}, {Fogantini}, {Chaty}  \&
  {Combi}}{{Garc{\'\i}a} et~al.}{2018}]{2018A&A...618A..61G}
{Garc{\'\i}a} F.,  {Fogantini} F.~A.,  {Chaty} S.,   {Combi} J.~A.,  2018,
  \mn@doi [\aap] {10.1051/0004-6361/201833365}, \href
  {https://ui.adsabs.harvard.edu/abs/2018A&A...618A..61G} {618, A61}

\bibitem[\protect\citeauthoryear{{Ghosh} \& {Lamb}}{{Ghosh} \&
  {Lamb}}{1979}]{1979ApJ...234..296G}
{Ghosh} P.,  {Lamb} F.~K.,  1979, \mn@doi [\apj] {10.1086/157498}, \href
  {https://ui.adsabs.harvard.edu/abs/1979ApJ...234..296G} {234, 296}

\bibitem[\protect\citeauthoryear{{Grebenev} \& {Sunyaev}}{{Grebenev} \&
  {Sunyaev}}{2007}]{2007AstL...33..149G}
{Grebenev} S.~A.,  {Sunyaev} R.~A.,  2007, \mn@doi [Astronomy Letters]
  {10.1134/S1063773707030024}, \href
  {https://ui.adsabs.harvard.edu/abs/2007AstL...33..149G} {33, 149}

\bibitem[\protect\citeauthoryear{{Harrison} et~al.,}{{Harrison}
  et~al.}{2013}]{2013ApJ...770..103H}
{Harrison} F.~A.,  et~al., 2013, \mn@doi [\apj] {10.1088/0004-637X/770/2/103},
  \href {https://ui.adsabs.harvard.edu/abs/2013ApJ...770..103H} {770, 103}

\bibitem[\protect\citeauthoryear{{Heemskerk} \& {van Paradijs}}{{Heemskerk} \&
  {van Paradijs}}{1989}]{1989A&A...223..154H}
{Heemskerk} M.~H.~M.,  {van Paradijs} J.,  1989, \aap, \href
  {https://ui.adsabs.harvard.edu/abs/1989A&A...223..154H} {223, 154}

\bibitem[\protect\citeauthoryear{{Hemphill}, {Rothschild}, {Markowitz},
  {F{\"u}rst}, {Pottschmidt}  \& {Wilms}}{{Hemphill}
  et~al.}{2014}]{2014ApJ...792...14H}
{Hemphill} P.~B.,  {Rothschild} R.~E.,  {Markowitz} A.,  {F{\"u}rst} F.,
  {Pottschmidt} K.,   {Wilms} J.,  2014, \mn@doi [\apj]
  {10.1088/0004-637X/792/1/14}, \href
  {https://ui.adsabs.harvard.edu/abs/2014ApJ...792...14H} {792, 14}

\bibitem[\protect\citeauthoryear{{Heras}, {Chaty}, {Prat}  \&
  {Rodriguez}}{{Heras} et~al.}{2009}]{2009AIPC.1126..313H}
{Heras} J.~A.~Z.,  {Chaty} S.,  {Prat} L.,   {Rodriguez} J.,  2009, in
  {Rodriguez} J.,  {Ferrando} P.,  eds,  American Institute of Physics
  Conference Series Vol. 1126, SIMBOL-X: Focusing on the Hard X-ray Universe.
  pp 313--315, \mn@doi{10.1063/1.3149440}

\bibitem[\protect\citeauthoryear{{Ho}, {Klus}, {Coe}  \& {Andersson}}{{Ho}
  et~al.}{2014}]{2014MNRAS.437.3664H}
{Ho} W. C.~G.,  {Klus} H.,  {Coe} M.~J.,   {Andersson} N.,  2014, \mn@doi
  [\mnras] {10.1093/mnras/stt2193}, \href
  {https://ui.adsabs.harvard.edu/abs/2014MNRAS.437.3664H} {437, 3664}

\bibitem[\protect\citeauthoryear{{Hutchings}, {Crampton}, {Cowley}  \&
  {Osmer}}{{Hutchings} et~al.}{1977}]{1977ApJ...217..186H}
{Hutchings} J.~B.,  {Crampton} D.,  {Cowley} A.~P.,   {Osmer} P.~S.,  1977,
  \mn@doi [\apj] {10.1086/155569}, \href
  {https://ui.adsabs.harvard.edu/abs/1977ApJ...217..186H} {217, 186}

\bibitem[\protect\citeauthoryear{{Jain}, {Paul}  \& {Dutta}}{{Jain}
  et~al.}{2009}]{2009MNRAS.397L..11J}
{Jain} C.,  {Paul} B.,   {Dutta} A.,  2009, \mn@doi [\mnras]
  {10.1111/j.1745-3933.2009.00668.x}, \href
  {https://ui.adsabs.harvard.edu/abs/2009MNRAS.397L..11J} {397, L11}

\bibitem[\protect\citeauthoryear{{Kretschmar} et~al.,}{{Kretschmar}
  et~al.}{2019}]{2019NewAR..8601546K}
{Kretschmar} P.,  et~al., 2019, \mn@doi [\nar] {10.1016/j.newar.2020.101546},
  \href {https://ui.adsabs.harvard.edu/abs/2019NewAR..8601546K} {86, 101546}

\bibitem[\protect\citeauthoryear{{Krimm} et~al.,}{{Krimm}
  et~al.}{2013}]{2013ApJS..209...14K}
{Krimm} H.~A.,  et~al., 2013, \mn@doi [\apjs] {10.1088/0067-0049/209/1/14},
  \href {https://ui.adsabs.harvard.edu/abs/2013ApJS..209...14K} {209, 14}

\bibitem[\protect\citeauthoryear{{Leahy}, {Darbro}, {Elsner}, {Weisskopf},
  {Sutherland}, {Kahn}  \& {Grindlay}}{{Leahy}
  et~al.}{1983}]{1983ApJ...266..160L}
{Leahy} D.~A.,  {Darbro} W.,  {Elsner} R.~F.,  {Weisskopf} M.~C.,  {Sutherland}
  P.~G.,  {Kahn} S.,   {Grindlay} J.~E.,  1983, \mn@doi [\apj]
  {10.1086/160766}, \href
  {https://ui.adsabs.harvard.edu/abs/1983ApJ...266..160L} {266, 160}

\bibitem[\protect\citeauthoryear{{Lutovinov}, {Rodriguez}, {Revnivtsev}  \&
  {Shtykovskiy}}{{Lutovinov} et~al.}{2005}]{2005A&A...433L..41L}
{Lutovinov} A.,  {Rodriguez} J.,  {Revnivtsev} M.,   {Shtykovskiy} P.,  2005,
  \mn@doi [\aap] {10.1051/0004-6361:200500092}, \href
  {https://ui.adsabs.harvard.edu/abs/2005A&A...433L..41L} {433, L41}

\bibitem[\protect\citeauthoryear{{Madsen} et~al.,}{{Madsen}
  et~al.}{2019}]{2019BAAS...51g.166M}
{Madsen} K.,  et~al., 2019, in Bulletin of the American Astronomical Society.
  p.~166

\bibitem[\protect\citeauthoryear{{Makishima}, {Mihara}, {Nagase}  \&
  {Tanaka}}{{Makishima} et~al.}{1999}]{makishima1999}
{Makishima} K.,  {Mihara} T.,  {Nagase} F.,   {Tanaka} Y.,  1999, \mn@doi
  [\apj] {10.1086/307912}, \href
  {http://adsabs.harvard.edu/abs/1999ApJ...525..978M} {525, 978}

\bibitem[\protect\citeauthoryear{{Miyasaka} et~al.,}{{Miyasaka}
  et~al.}{2013}]{2013ApJ...775...65M}
{Miyasaka} H.,  et~al., 2013, \mn@doi [\apj] {10.1088/0004-637X/775/1/65},
  \href {https://ui.adsabs.harvard.edu/abs/2013ApJ...775...65M} {775, 65}

\bibitem[\protect\citeauthoryear{{Molkov}, {Mowlavi}, {Goldwurm}, {Strong},
  {Lund}, {Paul}  \& {Oosterbroek}}{{Molkov}
  et~al.}{2003}]{2003ATel..176....1M}
{Molkov} S.,  {Mowlavi} N.,  {Goldwurm} A.,  {Strong} A.,  {Lund} N.,  {Paul}
  J.,   {Oosterbroek} T.,  2003, The Astronomer's Telegram, \href
  {https://ui.adsabs.harvard.edu/abs/2003ATel..176....1M} {176, 1}

\bibitem[\protect\citeauthoryear{{Nabizadeh}, {M{\"o}nkk{\"o}nen}, {Tsygankov},
  {Doroshenko}, {Molkov}  \& {Poutanen}}{{Nabizadeh}
  et~al.}{2019}]{2019A&A...629A.101N}
{Nabizadeh} A.,  {M{\"o}nkk{\"o}nen} J.,  {Tsygankov} S.~S.,  {Doroshenko} V.,
  {Molkov} S.~V.,   {Poutanen} J.,  2019, \mn@doi [\aap]
  {10.1051/0004-6361/201936045}, \href
  {https://ui.adsabs.harvard.edu/abs/2019A&A...629A.101N} {629, A101}

\bibitem[\protect\citeauthoryear{{Oskinova}, {Feldmeier}  \&
  {Kretschmar}}{{Oskinova} et~al.}{2012}]{2012MNRAS.421.2820O}
{Oskinova} L.~M.,  {Feldmeier} A.,   {Kretschmar} P.,  2012, \mn@doi [\mnras]
  {10.1111/j.1365-2966.2012.20507.x}, \href
  {https://ui.adsabs.harvard.edu/abs/2012MNRAS.421.2820O} {421, 2820}

\bibitem[\protect\citeauthoryear{{Postnov}, {Shakura}, {Kochetkova}  \&
  {Hjalmarsdotter}}{{Postnov} et~al.}{2012}]{2012int..workE..22P}
{Postnov} K.,  {Shakura} N.~I.,  {Kochetkova} A.~Y.,   {Hjalmarsdotter} L.,
  2012, in Proceedings of ``An INTEGRAL view of the high-energy sky (the first
  10 years)'' - 9th INTEGRAL Workshop and celebration of the 10th anniversary
  of the launch (INTEGRAL 2012). 15-19 October 2012. Bibliotheque Nationale de
  France. p.~22 (\mn@eprint {arXiv} {1212.2841}), \mn@doi{10.22323/1.176.0022}

\bibitem[\protect\citeauthoryear{{Pradhan}, {Bozzo}  \& {Paul}}{{Pradhan}
  et~al.}{2018}]{2018A&A...610A..50P}
{Pradhan} P.,  {Bozzo} E.,   {Paul} B.,  2018, \mn@doi [\aap]
  {10.1051/0004-6361/201731487}, \href
  {https://ui.adsabs.harvard.edu/abs/2018A&A...610A..50P} {610, A50}

\bibitem[\protect\citeauthoryear{{Puls}, {Vink}  \& {Najarro}}{{Puls}
  et~al.}{2008}]{2008A&ARv..16..209P}
{Puls} J.,  {Vink} J.~S.,   {Najarro} F.,  2008, \mn@doi [\aapr]
  {10.1007/s00159-008-0015-8}, \href
  {https://ui.adsabs.harvard.edu/abs/2008A&ARv..16..209P} {16, 209}

\bibitem[\protect\citeauthoryear{{Rodriguez}, {Tomsick}, {Foschini}, {Walter},
  {Goldwurm}, {Corbel}  \& {Kaaret}}{{Rodriguez}
  et~al.}{2003}]{2003A&A...407L..41R}
{Rodriguez} J.,  {Tomsick} J.~A.,  {Foschini} L.,  {Walter} R.,  {Goldwurm} A.,
   {Corbel} S.,   {Kaaret} P.,  2003, \mn@doi [\aap]
  {10.1051/0004-6361:20031093}, \href
  {https://ui.adsabs.harvard.edu/abs/2003A&A...407L..41R} {407, L41}

\bibitem[\protect\citeauthoryear{{Rodriguez} et~al.,}{{Rodriguez}
  et~al.}{2006}]{2006MNRAS.366..274R}
{Rodriguez} J.,  et~al., 2006, \mn@doi [\mnras]
  {10.1111/j.1365-2966.2005.09855.x}, \href
  {https://ui.adsabs.harvard.edu/abs/2006MNRAS.366..274R} {366, 274}

\bibitem[\protect\citeauthoryear{{Romano}}{{Romano}}{2015}]{2015JHEAp...7..126R}
{Romano} P.,  2015, \mn@doi [Journal of High Energy Astrophysics]
  {10.1016/j.jheap.2015.04.008}, \href
  {https://ui.adsabs.harvard.edu/abs/2015JHEAp...7..126R} {7, 126}

\bibitem[\protect\citeauthoryear{{Romano} et~al.,}{{Romano}
  et~al.}{2022}]{2022arXiv221205083R}
{Romano} P.,  et~al., 2022, arXiv e-prints, \href
  {https://ui.adsabs.harvard.edu/abs/2022arXiv221205083R} {p. arXiv:2212.05083}

\bibitem[\protect\citeauthoryear{{Sch{\"o}nherr}, {Wilms}, {Kretschmar},
  {Kreykenbohm}, {Santangelo}, {Rothschild}, {Coburn}  \&
  {Staubert}}{{Sch{\"o}nherr} et~al.}{2007}]{2007A&A...472..353S}
{Sch{\"o}nherr} G.,  {Wilms} J.,  {Kretschmar} P.,  {Kreykenbohm} I.,
  {Santangelo} A.,  {Rothschild} R.~E.,  {Coburn} W.,   {Staubert} R.,  2007,
  \mn@doi [\aap] {10.1051/0004-6361:20077218}, \href
  {https://ui.adsabs.harvard.edu/abs/2007A&A...472..353S} {472, 353}

\bibitem[\protect\citeauthoryear{{Sguera} et~al.,}{{Sguera}
  et~al.}{2005}]{2005A&A...444..221S}
{Sguera} V.,  et~al., 2005, \mn@doi [\aap] {10.1051/0004-6361:20053103}, \href
  {https://ui.adsabs.harvard.edu/abs/2005A&A...444..221S} {444, 221}

\bibitem[\protect\citeauthoryear{{Sguera} et~al.,}{{Sguera}
  et~al.}{2006}]{2006ApJ...646..452S}
{Sguera} V.,  et~al., 2006, \mn@doi [\apj] {10.1086/504827}, \href
  {https://ui.adsabs.harvard.edu/abs/2006ApJ...646..452S} {646, 452}

\bibitem[\protect\citeauthoryear{{Sguera} et~al.,}{{Sguera}
  et~al.}{2008}]{2008A&A...487..619S}
{Sguera} V.,  et~al., 2008, \mn@doi [\aap] {10.1051/0004-6361:20079195}, \href
  {https://ui.adsabs.harvard.edu/abs/2008A&A...487..619S} {487, 619}

\bibitem[\protect\citeauthoryear{{Sguera}, {Tiengo}, {Sidoli}  \&
  {Bird}}{{Sguera} et~al.}{2020}]{2020ApJ...900...22S}
{Sguera} V.,  {Tiengo} A.,  {Sidoli} L.,   {Bird} A.~J.,  2020, \mn@doi [\apj]
  {10.3847/1538-4357/abaa3c}, \href
  {https://ui.adsabs.harvard.edu/abs/2020ApJ...900...22S} {900, 22}

\bibitem[\protect\citeauthoryear{{Shakura}, {Postnov}, {Kochetkova}  \&
  {Hjalmarsdotter}}{{Shakura} et~al.}{2012}]{2012MNRAS.420..216S}
{Shakura} N.,  {Postnov} K.,  {Kochetkova} A.,   {Hjalmarsdotter} L.,  2012,
  \mn@doi [\mnras] {10.1111/j.1365-2966.2011.20026.x}, \href
  {https://ui.adsabs.harvard.edu/abs/2012MNRAS.420..216S} {420, 216}

\bibitem[\protect\citeauthoryear{{Shakura}, {Postnov}, {Sidoli}  \&
  {Paizis}}{{Shakura} et~al.}{2014}]{2014MNRAS.442.2325S}
{Shakura} N.,  {Postnov} K.,  {Sidoli} L.,   {Paizis} A.,  2014, \mn@doi
  [\mnras] {10.1093/mnras/stu1027}, \href
  {https://ui.adsabs.harvard.edu/abs/2014MNRAS.442.2325S} {442, 2325}

\bibitem[\protect\citeauthoryear{{Sidoli} \& {Paizis}}{{Sidoli} \&
  {Paizis}}{2019}]{2019IAUS..346..178S}
{Sidoli} L.,  {Paizis} A.,  2019, \mn@doi [IAU Symposium]
  {10.1017/S1743921319001145}, \href
  {https://ui.adsabs.harvard.edu/abs/2019IAUS..346..178S} {346, 178}

\bibitem[\protect\citeauthoryear{{Sidoli} et~al.,}{{Sidoli}
  et~al.}{2013}]{2013MNRAS.429.2763S}
{Sidoli} L.,  et~al., 2013, \mn@doi [\mnras] {10.1093/mnras/sts559}, \href
  {https://ui.adsabs.harvard.edu/abs/2013MNRAS.429.2763S} {429, 2763}

\bibitem[\protect\citeauthoryear{{Sundqvist}, {Owocki}  \& {Puls}}{{Sundqvist}
  et~al.}{2018}]{2018A&A...611A..17S}
{Sundqvist} J.~O.,  {Owocki} S.~P.,   {Puls} J.,  2018, \mn@doi [\aap]
  {10.1051/0004-6361/201731718}, \href
  {https://ui.adsabs.harvard.edu/abs/2018A&A...611A..17S} {611, A17}

\bibitem[\protect\citeauthoryear{{Sunyaev} \& {Titarchuk}}{{Sunyaev} \&
  {Titarchuk}}{1980}]{1980A&A....86..121S}
{Sunyaev} R.~A.,  {Titarchuk} L.~G.,  1980, \aap, \href
  {https://ui.adsabs.harvard.edu/abs/1980A&A....86..121S} {86, 121}

\bibitem[\protect\citeauthoryear{{Tanaka}}{{Tanaka}}{1986}]{T86}
{Tanaka} Y.,  1986, in {Mihalas} D.,  {Winkler} K.-H.~A.,  eds,  Lecture Notes
  in Physics, Berlin Springer Verlag Vol. 255, IAU Colloq. 89: Radiation
  Hydrodynamics in Stars and Compact Objects. p.~198,
  \mn@doi{10.1007/3-540-16764-1_12}

\bibitem[\protect\citeauthoryear{{Titarchuk}}{{Titarchuk}}{1994}]{1994ApJ...434..570T}
{Titarchuk} L.,  1994, \mn@doi [\apj] {10.1086/174760}, \href
  {https://ui.adsabs.harvard.edu/abs/1994ApJ...434..570T} {434, 570}

\bibitem[\protect\citeauthoryear{{Tomsick}, {Lingenfelter}, {Walter},
  {Rodriguez}, {Goldwurm}, {Corbel}  \& {Kaaret}}{{Tomsick}
  et~al.}{2003}]{2003IAUC.8076....1T}
{Tomsick} J.~A.,  {Lingenfelter} R.,  {Walter} R.,  {Rodriguez} J.,  {Goldwurm}
  A.,  {Corbel} S.,   {Kaaret} P.,  2003, \iaucirc, \href
  {https://ui.adsabs.harvard.edu/abs/2003IAUC.8076....1T} {8076, 1}

\bibitem[\protect\citeauthoryear{{VanderPlas} \& {Ivezi{\'c}}}{{VanderPlas} \&
  {Ivezi{\'c}}}{2015}]{2015ApJ...812...18V}
{VanderPlas} J.~T.,  {Ivezi{\'c}} {\v{Z}}.,  2015, \mn@doi [\apj]
  {10.1088/0004-637X/812/1/18}, \href
  {https://ui.adsabs.harvard.edu/abs/2015ApJ...812...18V} {812, 18}

\bibitem[\protect\citeauthoryear{{VanderPlas}, {Connolly}, {Ivezic}  \&
  {Gray}}{{VanderPlas} et~al.}{2012}]{2012cidu.conf...47V}
{VanderPlas} J.,  {Connolly} A.~J.,  {Ivezic} Z.,   {Gray} A.,  2012, in
  Proceedings of Conference on Intelligent Data Understanding (CIDU. pp 47--54
  (\mn@eprint {arXiv} {1411.5039}), \mn@doi{10.1109/CIDU.2012.6382200}

\bibitem[\protect\citeauthoryear{{Varun}, {Pradhan}, {Maitra}, {Raichur}  \&
  {Paul}}{{Varun} et~al.}{2019}]{2019ApJ...880...61V}
{Varun} {Pradhan} P.,  {Maitra} C.,  {Raichur} H.,   {Paul} B.,  2019, \mn@doi
  [\apj] {10.3847/1538-4357/ab2763}, \href
  {https://ui.adsabs.harvard.edu/abs/2019ApJ...880...61V} {880, 61}

\bibitem[\protect\citeauthoryear{{Varun}, {Iyer}  \& {Paul}}{{Varun}
  et~al.}{2023}]{2023NewA...9801942V}
{Varun} {Iyer} N.,   {Paul} B.,  2023, \mn@doi [\na]
  {10.1016/j.newast.2022.101942}, \href
  {https://ui.adsabs.harvard.edu/abs/2023NewA...9801942V} {98, 101942}

\bibitem[\protect\citeauthoryear{{Verner}, {Ferland}, {Korista}  \&
  {Yakovlev}}{{Verner} et~al.}{1996}]{1996ApJ...465..487V}
{Verner} D.~A.,  {Ferland} G.~J.,  {Korista} K.~T.,   {Yakovlev} D.~G.,  1996,
  \mn@doi [\apj] {10.1086/177435}, \href
  {https://ui.adsabs.harvard.edu/abs/1996ApJ...465..487V} {465, 487}

\bibitem[\protect\citeauthoryear{{Walter} \& {Zurita Heras}}{{Walter} \&
  {Zurita Heras}}{2007}]{2007A&A...476..335W}
{Walter} R.,  {Zurita Heras} J.,  2007, \mn@doi [\aap]
  {10.1051/0004-6361:20078353}, \href
  {https://ui.adsabs.harvard.edu/abs/2007A&A...476..335W} {476, 335}

\bibitem[\protect\citeauthoryear{{Wilms}, {Allen}  \& {McCray}}{{Wilms}
  et~al.}{2000}]{2000ApJ...542..914W}
{Wilms} J.,  {Allen} A.,   {McCray} R.,  2000, \mn@doi [\apj] {10.1086/317016},
  \href {https://ui.adsabs.harvard.edu/abs/2000ApJ...542..914W} {542, 914}

\bibitem[\protect\citeauthoryear{{{\.Z}ycki}, {Done}  \& {Smith}}{{{\.Z}ycki}
  et~al.}{1999}]{1999MNRAS.309..561Z}
{{\.Z}ycki} P.~T.,  {Done} C.,   {Smith} D.~A.,  1999, \mn@doi [\mnras]
  {10.1046/j.1365-8711.1999.02885.x}, \href
  {https://ui.adsabs.harvard.edu/abs/1999MNRAS.309..561Z} {309, 561}

\bibitem[\protect\citeauthoryear{{in 't Zand}, {Ubertini}, {Capitanio}  \& {Del
  Santo}}{{in 't Zand} et~al.}{2003}]{2003IAUC.8077....2I}
{in 't Zand} J.~J.~M.,  {Ubertini} P.,  {Capitanio} F.,   {Del Santo} M.,
  2003, \iaucirc, \href {https://ui.adsabs.harvard.edu/abs/2003IAUC.8077....2I}
  {8077, 2}

\makeatother
\end{thebibliography}

\bsp	
\label{lastpage}
\end{document}